\documentclass[preprint,aps]{revtex4}
\usepackage{mathrsfs}
\usepackage{amsmath}
\usepackage{graphicx}
\usepackage{multirow}
\usepackage{makecell}
\usepackage{booktabs}
\usepackage{dcolumn}
\usepackage{bm}
\usepackage{xcolor}
\usepackage[utf8]{inputenc}
\usepackage[T1]{fontenc}
\usepackage{comment}

\newcommand{\cvs}{CsV$_3$Sb$_5$}
\newcommand{\avs}{AV$_3$Sb$_5$}
\newcommand{\ef}{$E_{\rm F}$}
\newcommand{\tcdw}{$T_{\rm CDW}$}

\newcommand{\g}{$\Gamma$}

\newcommand{\edt}[1]{\textcolor{black}{#1}}

\begin{document}


\title{Magnetic field-induced momentum-dependent symmetry breaking in a kagome superconductor}

\author{Jianwei Huang$^{1,2,*,\dagger}$, Zheng Ren$^{1,2,*}$, Hengxin Tan$^{3}$, \edt{Jounghoon Hyun$^{1,2}$, Yichen Zhang$^{1,2}$, Thomas A. Hulse$^{1,2,4}$,} Zhaoyu Liu$^{5}$, \edt{Jonathan M. DeStefano$^{5}$, Yaofeng Xie$^{1,2}$}, Ziqin Yue$^{1,2,4}$, Junichiro Kono$^{1,2,6,7}$, \edt{Pengcheng Dai$^{1,8}$, Yu He$^{9}$, Aki Pulkkinen$^{10}$, Ján Minár$^{10}$}, Jiun-Haw Chu$^{5}$, Ziqiang Wang$^{11}$, Binghai Yan$^{3,12}$, Rafael M. Fernandes$^{13,14}$, Ming Yi$^{1,2,8\dagger}$}
\affiliation{
\\$^{1}$Department of Physics and Astronomy, Rice University, Houston, TX 77005, USA
\\$^{2}$Smalley–Curl Institute, Rice University, Houston, TX 77005, USA
\\$^{3}$Department of Condensed Matter Physics, Weizmann Institute of Science, Rehovot 7610001, Israel
\\$^{4}$Applied Physics Graduate Program, Smalley-Curl Institute, Rice University, Houston, TX 77005, USA
\\$^{5}$Department of Physics, University of Washington, Seattle, WA 98195, USA
\\$^{6}$Department of Electrical and Computer Engineering, Rice University, Houston, TX 77005, USA
\\$^{7}$Department of Materials Science and NanoEngineering, Rice University, Houston, TX 77005, USA
\\$^{8}$Rice Laboratory for Emergent Magnetic Materials, Rice University, Houston, TX 77005, USA
\\$^{9}$Department of Applied Physics, Yale University, New Haven, CT, 06511, USA
\\$^{10}$New Technologies - Research Centre, University of West Bohemia in Pilsen, Plzeň, 30100, Czech Republic
\\$^{11}$Department of Physics, Boston College, Chestnut Hill, Massachusetts 02467, USA
\\$^{12}$Department of Physics, The Pennsylvania State University, University Park, Pennsylvania 16802, USA
\\$^{13}$Department of Physics, University of Illinois Urbana-Champaign, Urbana, Illinois 61801, USA
\\$^{14}$Anthony J. Leggett Institute for Condensed Matter Theory, University of Illinois Urbana-Champaign, Urbana, Illinois 61801, USA
\\$^{*}$ These authors contributed equally 
\\$^{\dagger}$ To whom correspondence should be addressed: 
\\JH: jwhuang0rice@gmail.com
MY: mingyi@rice.edu
}

\date{\today}

\begin{abstract}
When multiple degrees of freedom share similar energy scales in quantum materials, intertwined electronic orders, which exhibit broken symmetries, are often strongly coupled. Recent studies on kagome superconductors such as \cvs~report rotational and time-reversal symmetry breaking linked to a charge density wave. Here, we observe a momentum-selective response of the electronic structure of \cvs~to an external magnetic field. By performing angle-resolved photoemission spectroscopy in a tuneable magnetic field, we demonstrate that the response of the electronic structure is compatible with piezomagnetism along with strong orbital selectivity. Our results show that the origin of the time-reversal symmetry breaking is associated with the vanadium Van Hove singularities at the onset of the charge density wave order.  We also demonstrate the presence of fluctuations beyond the charge ordering temperature. Our results reveal that magnetic fields can be used as tuning knobs for disentangling intertwined orders in the momentum space for quantum materials.

\end{abstract}

\maketitle

\newpage

\section{Main}

The kagome lattice, a prototypical two-dimensional geometrically frustrated system, has been extensively studied as a promising candidate for realizing the quantum spin liquid state~\cite{Syozi1951, Balents2010a, Han2012}. Beyond spin interactions, the kagome lattice also exhibits rich physics associated with the quantum interference of the electronic hopping paths, hosting flat bands, Dirac cones, and Van Hove singularities (VHS) in its electronic structure~\cite{Mielke1991, Tanaka2003, Yu2012a, Kiesel2012, Kiesel2013}. When the flat bands or the VHSs are positioned near the Fermi level, electronic instabilities emerge even for relatively weak electron-electron interactions, potentially leading to the emergence of different symmetry-broken phases, such as magnetism, charge density wave (CDW), superconductivity, and electronic nematicity~\cite{Norman2016, Ko2009, Yu2012a, Kiesel2012, Kiesel2013, Wang2013a}. Such a proximity to competing ordered states makes the kagome systems highly susceptible to external perturbations such as strain and magnetic field, which can be used to tune the competition among the exotic phases~\cite{Liu2020a, Wang2023a, Lima2023, Dey2011}. Consequently, kagome metals provide an excellent platform for investigating the various intriguing symmetry-breaking phases due to the intricate interweaving among electron correlations, topology, and geometric frustration~\cite{Balents2010a, Ko2009, Kang2020a}. 

\cvs~has been identified as a kagome superconductor (T$_c$ $\sim$ 3.2 K) with the V atoms forming the kagome lattice~\cite{Ortiz2019, Ortiz2020}. It undergoes a 2$\times$2$\times$2 CDW transition at 94 K, which is well captured by first-principles calculations, indicating the relevant role of the electron-phonon coupling~\cite{Tan2021,Christensen2021}. 
Interestingly, the CDW phase was also found to display unconventional properties suggestive of time-reversal symmetry-breaking (TRSB), which has been interpreted in terms of an interaction-driven loop-current phase \cite{Park2021,Lin2021,Denner2021,Christensen2022,Fischer2023,Tazai2024}. 
These properties include an enhanced relaxation rate measured by muon spin rotation~\cite{Mielke2022b, Khasanov2022}, magnetic-field switching of the relative intensity of the CDW Bragg peaks seen in scanning tunneling microscopy~\cite{Jiang2021,Xing2024}, and anisotropic and non-reciprocal electrical magneto-transport~\cite{Guo2022, Wei2024, Guo2024a}. On the other hand, the presence of a non-zero magnetization remains under debate, in view of contradictory reports on the existence of a spontaneous Kerr effect~\cite{Xu2022,Farhang2023, Saykin2023, Wang2024}. Additionally, sixfold (C$_6$) to twofold (C$_2$) rotational symmetry breaking (RSB) was detected at around 35 K in the CDW state of \cvs~by a combination of nuclear magnetic resonance and elastoresistance measurements~\cite{Nie2022a}. Such a RSB was attributed to the interlayer stacking of the 2$\times$2$\times$2 CDW order \cite{Christensen2021} and interpreted as an electronic nematic order\cite{Xiang2021, Jin2024}. While the interpretation of RSB in terms of nematic order has been challenged by recent studies that did not find a divergent nematic susceptibility~\cite{Liu2024}, various experimental probes in \cvs, including polarization-resolved Raman spectroscopy~\cite{Wulferding2022}, scanning tunneling microscopy~\cite{Zhao2021a, Chen2021a, Li2022a}, and optical spectroscopy \cite{Xu2022}, have consistently found RSB within the CDW state. Interestingly, resistivity anisotropy \cite{Guo2024a} and magnetic torque measurements ~\cite{Asaba2024} in \cvs~have proposed a connection between RSB and TRSB. Thus, to advance the understanding of this material and shed light on these conflicting experimental reports, a direct momentum-resolved spectroscopy measurement in the presence of a magnetic field is desirable.

In light of this, we utilize our recently developed magneto-ARPES technique~\cite{Huang2023a} to directly measure the response of the momentum-resolved electronic structure of \cvs~to an out-of-plane tunable magnetic field. ARPES has been remarkably successful in directly probing the momentum-resolved electronic spectra of quantum materials, providing a way to disentangle intertwined order parameters that exhibit distinct momentum-dependent fingerprints~\cite{Damascelli2003, Sobota2021}. For the \avs~(A = K, Rb, Cs) compounds, ARPES studies have revealed the key electronic characteristics inherent to the kagome lattice, along with the band folding and CDW gap associated with the CDW transition~\cite{Liu2021, Luo2022, Kang2022, Hu2022a}. However, to date, no momentum-resolved electronic response to an external magnetic field has been reported using ARPES.

Magnetic field is commonly used as a tuning knob for other types of experiments, but it has been carefully avoided in ARPES measurements due to its unpredictable effects on the photoelectron trajectory after photoemission~\cite{Ryu2023a, Huang2023a}. Recently, we have developed a simple method for implementing an \textit{in situ} tunable out-of-plane magnetic field as the sample environment during ARPES measurements via a solenoid coil (Fig.~\ref{fig:Fig1}e and f), enabling magneto-ARPES--a term coined by an independent work developing an in-plane field for ARPES~\cite{Ryu2023a}. While the magnetic fields available in our setup are small, below 10 mT, they are larger than the internal magnetic field values reported by $\mu$SR in \avs~\cite{Mielke2022b, Khasanov2022}. Utilizing this capability, we performed magneto-ARPES measurements on \cvs~to probe a momentum-dependent electronic response to a magnetic field. Specifically, we observed spectral modifications to the dispersions and Fermi surfaces associated with the vanadium $d$-orbital VHS bands around the K/K' points of the BZ. Importantly, these spectral modifications break the C$_6$ symmetry of the dispersion and are odd in the magnetic field, consistent with piezomagnetism \cite{Xing2024}. Additionally, we observe an elliptical elongation of the antimony $p$-orbital electron pocket around the \g~point of the BZ in the presence of a magnetic field. In contrast to the field-induced anisotropic spectral changes observed in the vanadium $d$-bands, which disappear above \tcdw, the spectral changes around the \g~point persists above \tcdw, suggesting significant fluctuations. Our magneto-ARPES results unambiguously reveal time-reversal symmetry breaking in \cvs~interwoven with rotational symmetry breaking, and a qualitatively different magnetic-field response of the low-energy electronic states arising from the Sb $p$-orbitals and the V $d$-orbitals. These results not only provide strong constraints to theoretical delineation of the exotic CDW order parameter in \cvs, but also establish a new capability for disentangling momentum-dependent order parameters in the presence of a field. 

\section{Field response of the V 3$d$ VHS bands at K/K'}

\begin{figure}
\includegraphics[width=1\textwidth]{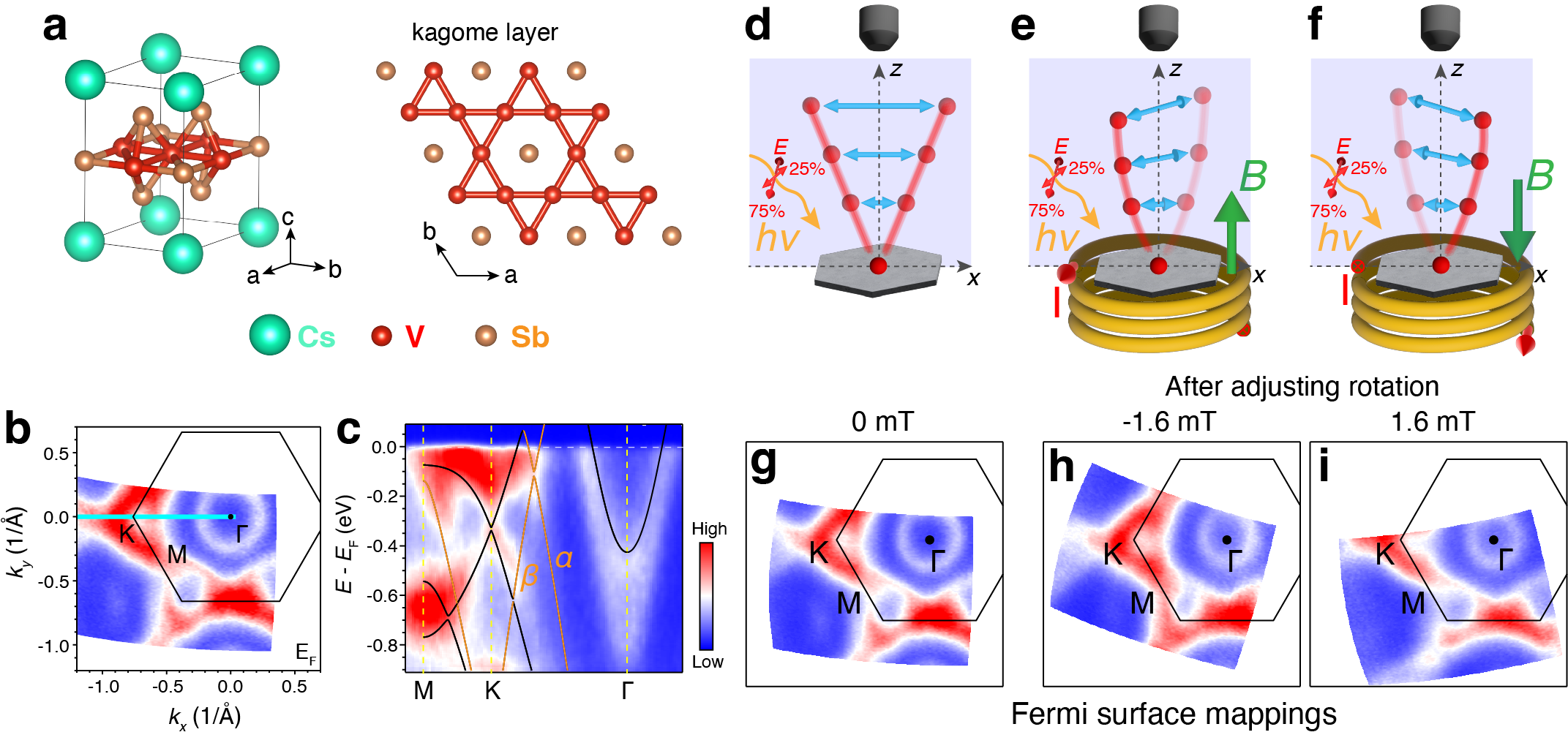}
\caption{\label{fig:Fig1}{\bf Magneto-ARPES experimental setup and the measured electronic structure of \cvs.} (a) Crystal structure of \cvs, where the V sublattice forms a kagome lattice. (b) Measured Fermi surface of \cvs~in the absence of a magnetic field, obtained by rotating the sample relative to the analyzer. (c) Spectral image extracted from the corresponding cut indicated in (b) overlaid with the DFT calculated band dispersions. The bands labeled $\alpha$ and $\beta$ respond most prominently to the external magnetic field, as revealed by magneto-ARPES. \edt{(d)-(f) Schematics of the magneto-ARPES experimental geometry and photoelectron trajectory with and without an \textit{in situ} magnetic field generated by a solenoid coil on the sample stage.} The magnetic field direction shown in (f) is defined as positive in this work. (g)-(i) The Fermi surfaces of \cvs~measured without and with a magnetic field illustrated in the configurations in (d)-(f). Fermi surface rotations are corrected post-measurement.
}
\end{figure}

We first introduce the zero-field electronic structure of \cvs~measured using a helium lamp-based ARPES system (Fig.~\ref{fig:Fig1}b and c). Consistent with previous reports, the Fermi surface of \cvs~is composed of a circular electron pocket around the BZ center consisting mainly of the 5$p$ orbitals of Sb and three triangles around the BZ corner (K/K') consisting of the 3$d$ orbitals of V and related to the nearby VHS at the M points ~\cite{Kang2022, Hu2022a}. The measured band dispersions match well with those obtained from density-functional theory (DFT) calculations (Fig.~\ref{fig:Fig1}c).
Raw magneto-ARPES data generally consist of intrinsic effects occurring inside the sample in response to the magnetic field and extrinsic effects occurring when the field acts on the photoemitted electrons on their way to the analyzer after they are ejected from the sample. The extrinsic effects consist of a constant energy contour rotation due to the Lorentz force, photoelectron emission angle contraction, and momentum broadening (Fig.~\ref{fig:Fig1}d-f)~\cite{Huang2023a}. Specifically, for the small magnetic field adopted in this work ($\sim$1.6 mT), only a constant energy contour rotation was discernible while the other two effects were negligible (Fig.~\ref{fig:Fig1}g and h)~\cite{Huang2023a}. \edt{This in-plane constant energy contour rotation is effectively energy independent within the energy range examined in this work (up to 1 eV below the Fermi level, see Supplementary Materials), allowing straightforward post-measurement correction without the need for intricate data processing.} It can also be used to directly determine the field strength at the sample. As an example, Fig.~\ref{fig:Fig1}g-i compares the Fermi surface mapped by rotating the sample measured at 0, -1.6 mT and +1.6 mT. The maps display opposite rotation directions for positive and negative fields, which can be corrected by an in-plane azimuthal angle offset before any further data analysis is performed (Fig.~\ref{fig:Fig1}h,i). To avoid any potential complications in field-dependent comparisons arising from the photoemission matrix-element effects~\cite{Damascelli2003}, all subsequent measurements were performed with the DA30 deflector mode without the need to rotate or move samples when changing the magnetic field. Hence, in all comparisons that we provide with and without a field, the beam spot was kept on a fixed spot of the sample during the whole measurement, guaranteeing that the light polarization relative to the crystal axes remained the same when the field strength was changed.  

We next compare the Fermi surface of \cvs~measured with and without a magnetic field in the CDW phase. To better observe the evolution of the electronic spectra in a magnetic field, we analyze the constant energy contours at -0.35 eV where the three sets of bands forming the triangular Fermi surfaces at the Fermi level (\ef) around K are well separated (Fig.~\ref{fig:Fig2}a). In the absence of a magnetic field, two large triangular pockets ($\alpha$ and $\beta$) and a small circular pocket at K (K') are resolved. The corresponding ARPES spectra are not distinguishable between K and K' (Fig.~\ref{fig:Fig2}b). Notably, the $\alpha$ and $\beta$ sheets meet at the M point, forming two sets of "X"-shaped constant energy contours that straddle the M point across the two neighboring BZs. When a magnetic field of 1.6 mT is applied, the overall shape of the constant energy contour remains unchanged. However, one diagonal branch of the "X"-shaped sheets weakens into the spectral background while the other branch remains sharp and discernible (yellow arrow in Fig.~\ref{fig:Fig2}c). Interestingly, when the direction of the magnetic field is reversed, the weakening behavior of the "X"-shaped constant energy contour also reverses between its two branches (yellow arrow in Fig.~\ref{fig:Fig2}d). 

To better illustrate this field-induced anisotropic spectral intensity change, $E-k$ spectral images along cuts across both K and K' are extracted for a side-by-side comparison (Fig.~\ref{fig:Fig2}e-h and Extended Data Fig. 1-2). The cuts are taken perpendicular to the $\Gamma$--M mirror plane, where a mirror-symmetric spectral weight distribution is expected when there is no magnetic field. \edt{We note that the experimental geometry is chosen to preserve this mirror, as neither our beam polarization nor incident direction breaks this mirror symmetry (Fig.~\ref{fig:Fig2})}. We focus on the $\alpha$ band as its spectral weight is discernible over a large energy range. For cut 1 within the first BZ, the ARPES spectra of the $\alpha$ band on the left side ($\alpha_{K'}$) and those of the $\alpha$ band on the right side ($\alpha_K$) of the $k_y$ = 0 mirror plane are equivalent in the absence of a magnetic field (Fig.~\ref{fig:Fig2}e). However, when an out-of-plane magnetic field (1.6 mT) is applied, the left branch of the $\alpha$ band intensity becomes "weak" and the spectral weight becomes broader, as observed from the momentum distribution curve (MDC) in Fig.~\ref{fig:Fig2}f. The spectral broadening is reversed with respect to the horizontal mirror plane when the magnetic field direction is reversed (Fig.~\ref{fig:Fig2}g). This can be better seen in the magneto-dichroic ARPES signal shown in Fig.~\ref{fig:Fig2}h, which we define as the difference in spectral intensity between the two spectra obtained in magnetic fields with opposite directions and the same strength. This magneto-dichroic signal can be well simulated by a simple selective momentum broadening of ARPES spectra based on the DFT calculated band structures along this cut (Extended Data Fig. 3). For cut 2, which lies outside the first BZ on the other side of the K--M--K BZ boundary, an opposite selective spectral weight broadening behavior is observed between $\alpha_{K'}$ and $\alpha_K$ in a magnetic field compared with what is observed along cut 1 (Extended Data Fig. 1). The spectral evolution of the $E-k$ images along both cut 1 and cut 2 in an external magnetic field together complements the spectral response of the "X"-shaped energy iso-surface to a magnetic field, which directly breaks the $\Gamma$--M mirror symmetry of the electronic states. We emphasize that this observation is an intrinsic electronic response of \cvs~to the external magnetic field, as the extrinsic momentum broadening effect is negligible at this small field and should be neither momentum selective nor dependent on the sign of the field~\cite{Huang2023a}.

To pinpoint the precise symmetries that are broken by this anisotropic effect, we note that our measurement region also covers part of the second BZ, which can be mapped back to the first BZ via the crystalline translational symmetry. We also directly verified this by rotating the sample and mapping the different K/K' pairs response to the magnetic field on the same sample (Extended Data Fig. 4). This leads to the illustration in Fig.~\ref{fig:Fig2}i, where the broadened branches of the Fermi surface are depicted by the thicker and lighter-colored lines while the shaper branches are represented by the thinner and darker-colored lines. Comparing the behavior of the branches across the first BZ, we conclude that the inversion symmetry is preserved, whereas the sixfold (C$_6$) rotational symmetry is broken. Moreover, since the effect changes sign when the field is reversed, the RSB is odd in field, which is indicative of piezomagnetism, i.e. a shear strain that is linear with the field \cite{Xing2024}.

To further explore this effect, we performed temperature-dependent measurements of the $E-k$ spectral images of \cvs~at a fixed magnetic field (Fig.~\ref{fig:Fig2}j-n and Supplementary Materials). Although the cut position (cut 3) slightly deviates from cut 1 in Fig.~\ref{fig:Fig2}b, it still captures both the $\alpha_{K'}$ and $\alpha_K$ bands (Fig.~\ref{fig:Fig2}j). As the temperature increases, the originally asymmetric $\alpha_{K'}$ and $\alpha_K$ bands in the CDW state restore their symmetry at higher temperatures. This is evident from the corresponding MDCs shown for $-0.42$ eV: while the two peaks have distinct widths at $20$ K, they gradually restore to similar widths at higher temperatures (Fig.~\ref{fig:Fig2}k). This evolution can also be visualized from the corresponding false-color map generated from the stacking of the temperature-dependent MDCs (Fig.~\ref{fig:Fig2}l). We can further quantify this temperature dependence by fitting both the peak height and peak width of $\alpha_{K'}$ and $\alpha_K$ from the MDCs. As seen from the resulting plots (Fig.~\ref{fig:Fig2}m-n), both the peak intensity and width become asymmetric below \tcdw, indicating that the symmetry breaking associated with the vanadium VHS bands is intrinsically linked to the CDW order.

Having demonstrated the magnetic-field-induced C$_6$ symmetry-breaking of the electronic spectrum in the CDW state of \cvs, we further examine its detailed spectral weight evolution as a function of magnetic field (Fig.~\ref{fig:Fig3} and Supplementary Materials). The $E-k$ spectral images along the $\Gamma$--K' direction measured at different magnetic fields are obtained as a cut from the Fermi surface mappings using the DA30 deflector mode after correcting for the Lorentz rotation (Fig.~\ref{fig:Fig3}a). In this measurement geometry, the aforementioned $\alpha$ band is suppressed whereas the $\beta$ band acquires a more clear resolution. As expected from the analysis of the $\alpha$ band, the spectral weight of the $\beta$ band significantly broadens at a positive magnetic field (2.1 mT), while the effect on the other bands are much less noticeable. To quantify this evolution, we extracted the MDCs at -0.2 eV from the $\Gamma$--K' spectral images (Fig.~\ref{fig:Fig3}b). Three distinct peaks (P1, P2, and P3) are resolved: P1 originates from the electron band around the BZ center, P3 from the shallow Dirac point at K', and P2 from the $\beta$ band. As the out-of-plane magnetic field is varied from negative to positive, peak P2 ($\beta$ band) broadens significantly in comparison to the linewidths of peak P1 and peak P3 (Fig.~\ref{fig:Fig3}d and top row of Fig.~\ref{fig:Fig3}a). \edt{The evolution on the peak area follows a similar trend, though varying less in comparison to the peak width.}

\edt{The odd-in-field response of the electronic structure to the out-of-plane field suggests signs of TRS-breaking. To understand the origin of the observed asymmetric spectral shape under a magnetic field, we carried out one-step model ARPES calculations using the spin-polarized relativistic Korringa-Kohn-Rostoker (SPRKKR) package to examine the effect of an external magnetic field on the photoemission spectral function on \cvs~\cite{Ebert2011Calculating} (see Methods and Extended Data Fig. 5). By calculating the spectral function of the non-CDW phase of \cvs~under an external field, we can simulate the photoemission spectra under a finite field for the simplest phase with broken TRS via the Zeeman effect without invoking a specific microscopic TRS-broken order. This set of calculations incorporates the magnetic field effects in all three parts of the photoemission matrix element—a Zeeman term for the initial state, the relativistic dipole operator, and the time-reversed spin-polarized low-energy electron diffraction final states, and is hence the simplest yet most unbiased way to examine the expectations from a generic TRS-broken state. As shown in Extended Data Fig. 5, the spectra near the K/K' points exhibit asymmetric spectral shape both in peak height and widths when photoemitting under a large magnetic field. This is due to the combination of the field-induced band shift and spin-dependent multiple scattering in the photoemission process under a particular field direction. While at first sight this may seem consistent with our observations, the magnitude of the field needed for the Zeeman effect to be observable is five orders of magnitude larger than the field used in the magneto-ARPES measurements. Hence the electronic response experimentally observed here cannot arise from a Zeeman-like field-induced effect on the photoemission matrix elements, but must originate from an intrinsic TRS-breaking order parameter in \cvs.
}

\section{Field response of the Sb 5$p$ band at \g}
Next, we turn our attention to the electron pocket around the BZ center (Fig.~\ref{fig:Fig4}), which arises from the Sb $p$ orbitals. When measured at zero field, as consistent with all previous reports~\cite{Liu2021, Luo2022, Kang2022, Hu2022a}, the central electron pocket is nearly isotropic and circular inside the CDW state (Fig.~\ref{fig:Fig4}a). This can be demonstrated by comparing the MDC along the vertical \g-K and horizontal \g-M directions, which shows that the separations of the two peaks are identical along the two orthogonal directions (Fig.~\ref{fig:Fig4}b). However, when an out-of-plane magnetic field of -1.6 mT is applied, the circular Fermi pocket becomes elliptical (Fig.~\ref{fig:Fig4}e), as can also be seen in the same comparison of the vertical (black) and horizontal (red) MDCs (Fig.~\ref{fig:Fig4}f). 
Here we note that we can exclude the possibility that this Fermi surface distortion is due to potential misalignment between the sample and the magnetic coil, \edt{ellipticity of the beam, or spatial inhomogeneity of the field}, as demonstrated via our extensive simulations (\edt{Supplementary Materials}). Rather, the elliptical distortion of the central Fermi pocket under the application of a magnetic field suggests a field-induced RSB in \cvs~corresponding to broken $C_6$ symmetry. 
Beyond the Fermi surface, we also show the measured dispersions along the vertical and horizontal directions for both zero field and -1.6 mT (Fig.~\ref{fig:Fig4}c,g). Their fitting shows that this elliptical distortion is not limited to the Fermi level but persists to a larger energy range of at least 250 meV below the Fermi level (Fig.~\ref{fig:Fig4}d,h). 

Another way to assess this anisotropy is to extract the fitted Fermi surface diameter as a function of the azimuthal angle ($\theta$). As shown in Fig.~\ref{fig:Fig4}i, the Fermi surface ellipticity for -1.6 mT is prominent compared to that at zero field at various temperatures. This trend is better seen by plotting the angle-dependent Fermi surface diameter in a magnetic field normalized by its zero-field value (Fig.~\ref{fig:Fig4}j). We note that this angle-dependent normalized Fermi surface diameter can be well fitted by a phenomenological function of the form $A + B\cos(2(\theta+\phi))$ that captures the two-fold rotational symmetry, with $B/A$ denoting the Fermi surface anisotropy ratio and the angle $\phi$ denoting the deviation of the long axis of the elliptical Fermi surface from the BZ high symmetry $\Gamma$--M ($\Gamma$--K) direction. Intriguingly, we find that in the CDW phase in a magnetic field, the long axis of the elliptical Fermi surface deviates from the high-symmetry $\Gamma$--K direction of the BZ (Fig.~\ref{fig:Fig4}j and k). At the same time, the directional deviation of the elliptical Fermi surface also changes with temperature, going through zero at \tcdw~and reversing sign above \tcdw~(Fig.~\ref{fig:Fig4}k). Additionally, the amplitude of the ellipticity of the Fermi surface remains relatively constant below \tcdw~and increases with increasing temperature above \tcdw~(Fig.~\ref{fig:Fig4}l). We plot the temperature dependence of the \g~pocket represented by the Fermi surface distortion together with the VHS spectral width difference extracted previously (Fig.~\ref{fig:Fig4}l). Their behaviors are distinct across \tcdw--the V 3$d$-VHS spectral width difference onsets below \tcdw~while the \g~electron pocket ellipticity is finite and largely constant below \tcdw, and grows above \tcdw~up to the measured temperature of 125 K (Fig.~\ref{fig:Fig4}m). 

\section{Discussion}
Our magneto-ARPES measurements reveal distinct field responses, inside the CDW phase, of the V 3$d$ orbitals that make up the bands associated with the VHS of the kagome lattice and the Sb $p$ orbitals that make up the band at the BZ center. The odd-in-field response of the VHS-bands breaks C$_6$ rotational symmetry, thus implying that the CDW phase of \cvs~is piezomagnetic, i.e., that an out-of-plane magnetic field induces an in-plane distortion. Piezomagnetism, manifesting the intrinsic coupling between TRSB and RSB, was also inferred in a recent STM study on the related compound RbV$_3$Sb$_5$~\cite{Xing2024}, which attributed this response to a so-called congruent flux phase with coexisting CDW and loop-current order. In this state, $C_6$ rotational symmetry is weakly broken already in the absence of a magnetic field, and the application of a magnetic field further breaks the remaining vertical mirror, leaving intact only the horizontal mirror. Our ARPES data at zero field, however, is $C_6$ symmetric, and the current magneto-ARPES data cannot resolve whether a vertical mirror plane is retained or not in the presence of a field. As discussed in the Methods section and in Extended Data Fig. 6-7, this could be due to the weak character of the zero-field $C_6$ symmetry-breaking or \edt{more likely domain averaging within the beamspot of the ARPES measurement. As our samples are glued down, they are not free from residual strain. Considering that our beamspot is much larger than the congruent flux order domains, in the absence of field, the time-reversed pairs of loop current domains would both appear, averaging out in our ARPES signal. With the application of a field and also in the presence of residual strain, one type of time-reversed loop current domain would dominate, revealing the observed electronic response.} To distinguish these scenarios from the one where $C_6$ symmetry is preserved at zero-field, future \edt{nano-focused} synchrotron-based magneto-ARPES measurements will be helpful. Regardless of its microscopic origin, the piezomagnetic response, which implies that the CDW state spontaneously breaks TRS, is robust and strongly manifested in the V $d$ bands that give rise to the VHS, which are often associated with the emergence of unconventional correlated states in the kagome lattice \cite{Kiesel2013}. A field-induced breaking of $C_6$ symmetry is also manifested in the bands made up of the Sb $p$ orbitals, which have been proposed to be essential for superconductivity \cite{Ritz2023_SC}. Our measurements on this Fermi pocket ellipticity around the BZ center, however, do not exhibit a clear sign change upon reversing the magnetic field (Extended Data Fig. 8), in contrast to the clear odd-in-field behavior displayed by the anisotropy of the V $d$ bands.

It is interesting to note that the field-response of the V 3$d$ bands vanishes at \tcdw~while that of the Sb $p_z$ bands persists above \tcdw. The origin of the latter behavior and its connection to the CDW order will require further studies. Interestingly, a recent magneto-torque measurement reported rotational symmetry breaking above \tcdw~that was interpreted to also break inversion symmetry \cite{Asaba2024}, which is not the case in our magneto-ARPES data.  
One possibility is that strong fluctuations persist above \tcdw~and are thus ``picked up'' by the magnetic field, similarly to how small residual strain induces robust nematic order in the iron-based superconductors due to an enlarged nematic susceptibility \cite{Fernandes2014a}. Indeed, it has been demonstrated that \cvs~is extremely sensitive to both magnetic field and strain~\cite{Guo2024}, and our glued sample is not free from random residual strain, which may combine with the external magnetic field to induce an ordered state. Alternatively, the magnetic field could also enhance the separation between the onsets of the CDW and loop-current order parameters in certain loop-current configurations~\cite{Christensen2022,Fischer2023,Tazai2024}. In either scenario, however, it is puzzling why this behavior is seen only in the Sb bands and why the anisotropy is larger outside the CDW phase than inside the CDW phase.  
It remains to be seen whether future experiments can map the electronic response of the distinct momentum-space regions to decoupled strain and magnetic fields, which will be helpful in disentangling different microscopic mechanisms. The new capability of magneto-ARPES demonstrated here, which enables the mapping of the spectral response in a momentum-resolved way in the presence of a tunable finite magnetic field, opens up that possibility.

\bibliographystyle{naturemag}
\bibliography{CVS_MARPES}

@article{Ebert2011Calculating,
  author    = {Ebert, H. and Ködderitzsch, D. and Minár, J.},
  title     = {Calculating condensed matter properties using the KKR-Green’s function method - Recent developments and applications},
  journal   = {Reports on Progress in Physics},
  volume    = {74},
  pages     = {096501},
  year      = {2011},
  publisher = {IOP Publishing},
  doi       = {10.1088/0034-4885/74/9/096501},
  url       = {https://iopscience.iop.org/article/10.1088/0034-4885/74/9/096501}
}

@article{Vosko1980Accurate,
  author    = {Vosko, S. H. and Wilk, L. and Nusair, M.},
  title     = {Accurate spin-dependent electron liquid correlation energies for local spin density calculations: a critical analysis},
  journal   = {Canadian Journal of Physics},
  volume    = {58},
  pages     = {1200--1211},
  year      = {1980},
  publisher = {Canadian Science Publishing},
  doi       = {10.1139/p80-159},
  url       = {https://cdnsciencepub.com/doi/10.1139/p80-159}
}

@article{Rundgren1977Transmission,
  author    = {Rundgren, J. and Malmstrom, J.},
  title     = {Transmission and reflection of low-energy electrons at the surface barrier of a metal},
  journal   = {Journal of Physics C: Solid State Physics},
  volume    = {10},
  pages     = {4671},
  year      = {1977},
  publisher = {IOP Publishing},
  doi       = {10.1088/0022-3719/10/23/004},
  url       = {https://iopscience.iop.org/article/10.1088/0022-3719/10/23/004}
}

@article{Braun1996Theory,
  author    = {Braun, J.},
  title     = {The theory of angle-resolved ultraviolet photoemission and its applications to ordered materials},
  journal   = {Reports on Progress in Physics},
  volume    = {59},
  pages     = {1267},
  year      = {1996},
  publisher = {IOP Publishing},
  doi       = {10.1088/0034-4885/59/10/002},
  url       = {https://iopscience.iop.org/article/10.1088/0034-4885/59/10/002}
}

@article{Syozi1951,
author = {Syozi, I.},
doi = {10.1143/ptp/6.3.306},
issn = {0033-068X},
journal = {Progress of Theoretical Physics},
month = {jun},
number = {3},
pages = {306--308},
title = {{Statistics of Kagome Lattice}},
url = {https://academic.oup.com/ptp/article-lookup/doi/10.1143/ptp/6.3.306},
volume = {6},
year = {1951}
}

@article{Mielke1991,
author = {Mielke, A},
doi = {10.1088/0305-4470/24/14/018},
issn = {0305-4470},
journal = {Journal of Physics A: Mathematical and General},
month = {jul},
number = {14},
pages = {3311--3321},
title = {{Ferromagnetism in the Hubbard model on line graphs and further considerations}},
url = {https://iopscience.iop.org/article/10.1088/0305-4470/24/14/018},
volume = {24},
year = {1991}
}

@article{Kresse1996b,
author = {Kresse, G. and Furthm{\"{u}}ller, J.},
doi = {10.1103/PhysRevB.54.11169},
issn = {0163-1829},
journal = {Physical Review B},
month = {oct},
number = {16},
pages = {11169--11186},
title = {{Efficient iterative schemes for ab initio total-energy calculations using a plane-wave basis set}},
url = {https://link.aps.org/doi/10.1103/PhysRevB.54.11169},
volume = {54},
year = {1996}
}

@article{Perdew1997,
author = {Perdew, John P. and Burke, Kieron and Ernzerhof, Matthias},
doi = {10.1103/PhysRevLett.78.1396},
issn = {0031-9007},
journal = {Physical Review Letters},
month = {feb},
number = {7},
pages = {1396--1396},
title = {{Generalized Gradient Approximation Made Simple [Phys. Rev. Lett. 77, 3865 (1996)]}},
url = {https://link.aps.org/doi/10.1103/PhysRevLett.78.1396},
volume = {78},
year = {1997}
}

@article{Tanaka2003,
author = {Tanaka, Akinori and Ueda, Hiromitsu},
doi = {10.1103/PhysRevLett.90.067204},
issn = {0031-9007},
journal = {Physical Review Letters},
month = {feb},
number = {6},
pages = {067204},
title = {{Stability of Ferromagnetism in the Hubbard Model on the Kagome Lattice}},
url = {https://link.aps.org/doi/10.1103/PhysRevLett.90.067204},
volume = {90},
year = {2003}
}

@article{Damascelli2003,
author = {Damascelli, Andrea and Hussain, Zahid and Shen, Zhi-Xun},
doi = {10.1103/RevModPhys.75.473},
issn = {0034-6861},
journal = {Reviews of Modern Physics},
month = {apr},
number = {2},
pages = {473--541},
title = {{Angle-resolved photoemission studies of the cuprate superconductors}},
url = {https://link.aps.org/doi/10.1103/RevModPhys.75.473},
volume = {75},
year = {2003}
}

@article{Ko2009,
author = {Ko, Wing-Ho and Lee, Patrick A. and Wen, Xiao-Gang},
doi = {10.1103/PhysRevB.79.214502},
issn = {1098-0121},
journal = {Physical Review B},
month = {jun},
number = {21},
pages = {214502},
title = {{Doped kagome system as exotic superconductor}},
url = {https://link.aps.org/doi/10.1103/PhysRevB.79.214502},
volume = {79},
year = {2009}
}

@article{Balents2010a,
author = {Balents, Leon},
doi = {10.1038/nature08917},
issn = {0028-0836},
journal = {Nature},
month = {mar},
number = {7286},
pages = {199--208},
title = {{Spin liquids in frustrated magnets}},
url = {https://www.nature.com/articles/nature08917},
volume = {464},
year = {2010}
}

@article{Dey2011,
author = {Dey, Moumita and Maiti, Santanu K. and Karmakar, S. N.},
doi = {10.1063/1.3658253},
issn = {0021-8979},
journal = {Journal of Applied Physics},
month = {nov},
number = {9},
title = {{Magnetic field induced metal-insulator transition in a kagome nanoribbon}},
url = {https://pubs.aip.org/jap/article/110/9/094306/945365/Magnetic-field-induced-metal-insulator-transition},
volume = {110},
pages = {094306},
year = {2011}
}

@article{Kiesel2012,
author = {Kiesel, Maximilian L. and Thomale, Ronny},
doi = {10.1103/PhysRevB.86.121105},
issn = {1098-0121},
journal = {Physical Review B},
month = {sep},
number = {12},
pages = {121105},
title = {{Sublattice interference in the kagome Hubbard model}},
url = {https://link.aps.org/doi/10.1103/PhysRevB.86.121105},
volume = {86},
year = {2012}
}

@article{Yu2012a,
author = {Yu, Shun-Li and Li, Jian-Xin},
doi = {10.1103/PhysRevB.85.144402},
issn = {1098-0121},
journal = {Physical Review B},
month = {apr},
number = {14},
pages = {144402},
title = {{Chiral superconducting phase and chiral spin-density-wave phase in a Hubbard model on the kagome lattice}},
url = {https://link.aps.org/doi/10.1103/PhysRevB.85.144402},
volume = {85},
year = {2012}
}

@article{Han2012,
author = {Han, Tian-Heng and Helton, Joel S. and Chu, Shaoyan and Nocera, Daniel G. and Rodriguez-Rivera, Jose A. and Broholm, Collin and Lee, Young S.},
doi = {10.1038/nature11659},
issn = {0028-0836},
journal = {Nature},
month = {dec},
number = {7429},
pages = {406--410},
title = {{Fractionalized excitations in the spin-liquid state of a kagome-lattice antiferromagnet}},
url = {https://www.nature.com/articles/nature11659},
volume = {492},
year = {2012}
}

@article{Kiesel2013,
author = {Kiesel, Maximilian L. and Platt, Christian and Thomale, Ronny},
doi = {10.1103/PhysRevLett.110.126405},
issn = {0031-9007},
journal = {Physical Review Letters},
month = {mar},
number = {12},
pages = {126405},
title = {{Unconventional Fermi Surface Instabilities in the Kagome Hubbard Model}},
url = {https://link.aps.org/doi/10.1103/PhysRevLett.110.126405},
volume = {110},
year = {2013}
}

@article{Wang2013a,
author = {Wang, Wan-Sheng and Li, Zheng-Zhao and Xiang, Yuan-Yuan and Wang, Qiang-Hua},
doi = {10.1103/PhysRevB.87.115135},
issn = {1098-0121},
journal = {Physical Review B},
month = {mar},
number = {11},
pages = {115135},
title = {{Competing electronic orders on kagome lattices at van Hove filling}},
url = {https://link.aps.org/doi/10.1103/PhysRevB.87.115135},
volume = {87},
year = {2013}
}

@article{Fernandes2014a,
author = {Fernandes, R. M. and Chubukov, A. V. and Schmalian, J.},
doi = {10.1038/nphys2877},
issn = {1745-2473},
journal = {Nature Physics},
month = {feb},
number = {2},
pages = {97--104},
title = {{What drives nematic order in iron-based superconductors?}},
url = {https://www.nature.com/articles/nphys2877},
volume = {10},
year = {2014}
}

@article{Norman2016,
author = {Norman, M. R.},
doi = {10.1103/RevModPhys.88.041002},
issn = {0034-6861},
journal = {Reviews of Modern Physics},
month = {dec},
number = {4},
pages = {041002},
title = {{Colloquium : Herbertsmithite and the search for the quantum spin liquid}},
url = {https://link.aps.org/doi/10.1103/RevModPhys.88.041002},
volume = {88},
year = {2016}
}

@article{Ortiz2019,
author = {Ortiz, Brenden R. and Gomes, L{\'{i}}dia C. and Morey, Jennifer R. and Winiarski, Michal and Bordelon, Mitchell and Mangum, John S. and Oswald, Iain W. H. and Rodriguez-Rivera, Jose A. and Neilson, James R. and Wilson, Stephen D. and Ertekin, Elif and McQueen, Tyrel M. and Toberer, Eric S.},
doi = {10.1103/PhysRevMaterials.3.094407},
issn = {2475-9953},
journal = {Physical Review Materials},
month = {sep},
number = {9},
pages = {094407},
title = {{New kagome prototype materials: discovery of KV$_3$Sb$_5$, RbV$_3$Sb$_5$, and CsV$_3$Sb$_5$}},
url = {https://link.aps.org/doi/10.1103/PhysRevMaterials.3.094407},
volume = {3},
year = {2019}
}

@article{Liu2020a,
author = {Liu, Tianyu},
doi = {10.1103/PhysRevB.102.045151},
issn = {2469-9950},
journal = {Physical Review B},
month = {jul},
number = {4},
pages = {045151},
title = {{Strain-induced pseudomagnetic field and quantum oscillations in kagome crystals}},
url = {https://link.aps.org/doi/10.1103/PhysRevB.102.045151},
volume = {102},
year = {2020}
}

@article{Kang2020a,
archivePrefix = {arXiv},
arxivId = {1906.02167},
author = {Kang, Mingu and Ye, Linda and Fang, Shiang and You, Jhih Shih and Levitan, Abe and Han, Minyong and Facio, Jorge I. and Jozwiak, Chris and Bostwick, Aaron and Rotenberg, Eli and Chan, Mun K. and McDonald, Ross D. and Graf, David and Kaznatcheev, Konstantine and Vescovo, Elio and Bell, David C. and Kaxiras, Efthimios and van den Brink, Jeroen and Richter, Manuel and {Prasad Ghimire}, Madhav and Checkelsky, Joseph G. and Comin, Riccardo},
doi = {10.1038/s41563-019-0531-0},
eprint = {1906.02167},
issn = {14764660},
journal = {Nature Materials},
number = {2},
pages = {163--169},
pmid = {31819211},
publisher = {Springer US},
title = {{Dirac fermions and flat bands in the ideal kagome metal FeSn}},
url = {http://dx.doi.org/10.1038/s41563-019-0531-0},
volume = {19},
year = {2020}
}

@article{Ortiz2020,
author = {Ortiz, Brenden R. and Teicher, Samuel M. L. and Hu, Yong and Zuo, Julia L. and Sarte, Paul M. and Schueller, Emily C. and Abeykoon, A. M. Milinda and Krogstad, Matthew J. and Rosenkranz, Stephan and Osborn, Raymond and Seshadri, Ram and Balents, Leon and He, Junfeng and Wilson, Stephen D.},
doi = {10.1103/PhysRevLett.125.247002},
issn = {0031-9007},
journal = {Physical Review Letters},
month = {dec},
number = {24},
pages = {247002},
title = {{CsV$_3$Sb$_5$: a Z$_2$ topological kagome metal with a superconducting ground state.}},
url = {https://link.aps.org/doi/10.1103/PhysRevLett.125.247002},
volume = {125},
year = {2020}
}

@article{Liu2021,
author = {Liu, Zhonghao and Zhao, Ningning and Yin, Qiangwei and Gong, Chunsheng and Tu, Zhijun and Li, Man and Song, Wenhua and Liu, Zhengtai and Shen, Dawei and Huang, Yaobo and Liu, Kai and Lei, Hechang and Wang, Shancai},
doi = {10.1103/PhysRevX.11.041010},
issn = {2160-3308},
journal = {Physical Review X},
month = {oct},
number = {4},
pages = {041010},
title = {{Charge-Density-Wave-Induced Bands Renormalization and Energy Gaps in a Kagome Superconductor RbV$_3$Sb$_5$}},
url = {https://link.aps.org/doi/10.1103/PhysRevX.11.041010},
volume = {11},
year = {2021}
}

@article{Chen2021a,
author = {Chen, Hui and Yang, Haitao and Hu, Bin and Zhao, Zhen and Yuan, Jie and Xing, Yuqing and Qian, Guojian and Huang, Zihao and Li, Geng and Ye, Yuhan and Ma, Sheng and Ni, Shunli and Zhang, Hua and Yin, Qiangwei and Gong, Chunsheng and Tu, Zhijun and Lei, Hechang and Tan, Hengxin and Zhou, Sen and Shen, Chengmin and Dong, Xiaoli and Yan, Binghai and Wang, Ziqiang and Gao, Hong-Jun},
doi = {10.1038/s41586-021-03983-5},
issn = {0028-0836},
journal = {Nature},
month = {nov},
number = {7884},
pages = {222--228},
title = {{Roton pair density wave in a strong-coupling kagome superconductor}},
url = {https://www.nature.com/articles/s41586-021-03983-5},
volume = {599},
year = {2021}
}

@article{Zhao2021a,
author = {Zhao, He and Li, Hong and Ortiz, Brenden R. and Teicher, Samuel M. L. and Park, Takamori and Ye, Mengxing and Wang, Ziqiang and Balents, Leon and Wilson, Stephen D. and Zeljkovic, Ilija},
doi = {10.1038/s41586-021-03946-w},
issn = {0028-0836},
journal = {Nature},
month = {nov},
number = {7884},
pages = {216--221},
title = {{Cascade of correlated electron states in the kagome superconductor CsV$_3$Sb$_5$}},
url = {https://www.nature.com/articles/s41586-021-03946-w},
volume = {599},
year = {2021}
}

@article{Qian2021,
author = {Qian, Tiema and Christensen, Morten H. and Hu, Chaowei and Saha, Amartyajyoti and Andersen, Brian M. and Fernandes, Rafael M. and Birol, Turan and Ni, Ni},
doi = {10.1103/PhysRevB.104.144506},
issn = {2469-9950},
journal = {Physical Review B},
month = {oct},
number = {14},
pages = {144506},
title = {{Revealing the competition between charge density wave and superconductivity in CsV$_3$Sb$_5$ through uniaxial strain}},
url = {https://link.aps.org/doi/10.1103/PhysRevB.104.144506},
volume = {104},
year = {2021}
}

@article{Jiang2021,
author = {Jiang, Yu-Xiao and Yin, Jia-Xin and Denner, M. Michael and Shumiya, Nana and Ortiz, Brenden R. and Xu, Gang and Guguchia, Zurab and He, Junyi and Hossain, Md Shafayat and Liu, Xiaoxiong and Ruff, Jacob and Kautzsch, Linus and Zhang, Songtian S. and Chang, Guoqing and Belopolski, Ilya and Zhang, Qi and Cochran, Tyler A. and Multer, Daniel and Litskevich, Maksim and Cheng, Zi-Jia and Yang, Xian P. and Wang, Ziqiang and Thomale, Ronny and Neupert, Titus and Wilson, Stephen D. and Hasan, M. Zahid},
doi = {10.1038/s41563-021-01034-y},
issn = {1476-1122},
journal = {Nature Materials},
month = {oct},
number = {10},
pages = {1353--1357},
title = {{Unconventional chiral charge order in kagome superconductor KV$_3$Sb$_5$}},
url = {https://www.nature.com/articles/s41563-021-01034-y},
volume = {20},
year = {2021}
}

@article{Sobota2021,
author = {Sobota, Jonathan A. and He, Yu and Shen, Zhi-Xun},
doi = {10.1103/RevModPhys.93.025006},
issn = {0034-6861},
journal = {Reviews of Modern Physics},
month = {may},
number = {2},
pages = {025006},
title = {{Angle-resolved photoemission studies of quantum materials}},
url = {https://link.aps.org/doi/10.1103/RevModPhys.93.025006},
volume = {93},
year = {2021}
}

@article{Tan2021,
author = {Tan, Hengxin and Liu, Yizhou and Wang, Ziqiang and Yan, Binghai},
doi = {10.1103/PhysRevLett.127.046401},
issn = {0031-9007},
journal = {Physical Review Letters},
month = {jul},
number = {4},
pages = {046401},
title = {{Charge Density Waves and Electronic Properties of Superconducting Kagome Metals}},
url = {https://link.aps.org/doi/10.1103/PhysRevLett.127.046401},
volume = {127},
year = {2021}
}

@article{Luo2022,
author = {Luo, Hailan and Gao, Qiang and Liu, Hongxiong and Gu, Yuhao and Wu, Dingsong and Yi, Changjiang and Jia, Junjie and Wu, Shilong and Luo, Xiangyu and Xu, Yu and Zhao, Lin and Wang, Qingyan and Mao, Hanqing and Liu, Guodong and Zhu, Zhihai and Shi, Youguo and Jiang, Kun and Hu, Jiangping and Xu, Zuyan and Zhou, X. J.},
doi = {10.1038/s41467-021-27946-6},
issn = {2041-1723},
journal = {Nature Communications},
month = {jan},
number = {1},
pages = {273},
title = {{Electronic nature of charge density wave and electron-phonon coupling in kagome superconductor KV$_3$Sb$_5$}},
url = {https://www.nature.com/articles/s41467-021-27946-6},
volume = {13},
year = {2022}
}

@article{Li2022a,
author = {Li, Hong and Zhao, He and Ortiz, Brenden R. and Park, Takamori and Ye, Mengxing and Balents, Leon and Wang, Ziqiang and Wilson, Stephen D. and Zeljkovic, Ilija},
doi = {10.1038/s41567-021-01479-7},
issn = {1745-2473},
journal = {Nature Physics},
month = {mar},
number = {3},
pages = {265--270},
title = {{Rotation symmetry breaking in the normal state of a kagome superconductor KV$_3$Sb$_5$}},
url = {https://www.nature.com/articles/s41567-021-01479-7},
volume = {18},
year = {2022}
}

@article{Wulferding2022,
author = {Wulferding, Dirk and Lee, Seungyeol and Choi, Youngsu and Yin, Qiangwei and Tu, Zhijun and Gong, Chunsheng and Lei, Hechang and Yousuf, Saqlain and Song, Jaegu and Lee, Hanoh and Park, Tuson and Choi, Kwang-Yong},
doi = {10.1103/PhysRevResearch.4.023215},
issn = {2643-1564},
journal = {Physical Review Research},
month = {jun},
number = {2},
pages = {023215},
title = {{Emergent nematicity and intrinsic versus extrinsic electronic scattering processes in the kagome metal CsV$_3$Sb$_5$}},
url = {https://link.aps.org/doi/10.1103/PhysRevResearch.4.023215},
volume = {4},
year = {2022}
}

@article{Guo2022,
author = {Guo, Chunyu and Putzke, Carsten and Konyzheva, Sofia and Huang, Xiangwei and Gutierrez-Amigo, Martin and Errea, Ion and Chen, Dong and Vergniory, Maia G. and Felser, Claudia and Fischer, Mark H. and Neupert, Titus and Moll, Philip J. W.},
doi = {10.1038/s41586-022-05127-9},
issn = {0028-0836},
journal = {Nature},
month = {nov},
number = {7936},
pages = {461--466},
title = {{Switchable chiral transport in charge-ordered kagome metal CsV$_3$Sb$_5$}},
url = {https://www.nature.com/articles/s41586-022-05127-9},
volume = {611},
year = {2022}
}

@article{Kang2022,
author = {Kang, Mingu and Fang, Shiang and Kim, Jeong-Kyu and Ortiz, Brenden R. and Ryu, Sae Hee and Kim, Jimin and Yoo, Jonggyu and Sangiovanni, Giorgio and {Di Sante}, Domenico and Park, Byeong-Gyu and Jozwiak, Chris and Bostwick, Aaron and Rotenberg, Eli and Kaxiras, Efthimios and Wilson, Stephen D. and Park, Jae-Hoon and Comin, Riccardo},
doi = {10.1038/s41567-021-01451-5},
issn = {1745-2473},
journal = {Nature Physics},
month = {mar},
number = {3},
pages = {301--308},
title = {{Twofold van Hove singularity and origin of charge order in topological kagome superconductor CsV$_3$Sb$_5$}},
url = {https://www.nature.com/articles/s41567-021-01451-5},
volume = {18},
year = {2022}
}

@article{Xu2022,
author = {Xu, Yishuai and Ni, Zhuoliang and Liu, Yizhou and Ortiz, Brenden R. and Deng, Qinwen and Wilson, Stephen D. and Yan, Binghai and Balents, Leon and Wu, Liang},
doi = {10.1038/s41567-022-01805-7},
issn = {1745-2473},
journal = {Nature Physics},
month = {dec},
number = {12},
pages = {1470--1475},
title = {{Three-state nematicity and magneto-optical Kerr effect in the charge density waves in kagome superconductors}},
url = {https://www.nature.com/articles/s41567-022-01805-7},
volume = {18},
year = {2022}
}

@article{Mielke2022b,
author = {Mielke, C. and Das, D. and Yin, J.-X. and Liu, H. and Gupta, R. and Jiang, Y.-X. and Medarde, M. and Wu, X. and Lei, H. C. and Chang, J. and Dai, Pengcheng and Si, Q. and Miao, H. and Thomale, R. and Neupert, T. and Shi, Y. and Khasanov, R. and Hasan, M. Z. and Luetkens, H. and Guguchia, Z.},
doi = {10.1038/s41586-021-04327-z},
issn = {0028-0836},
journal = {Nature},
month = {feb},
number = {7896},
pages = {245--250},
title = {{Time-reversal symmetry-breaking charge order in a kagome superconductor}},
url = {https://www.nature.com/articles/s41586-021-04327-z},
volume = {602},
year = {2022}
}

@article{Hu2022a,
author = {Hu, Yong and Wu, Xianxin and Ortiz, Brenden R. and Ju, Sailong and Han, Xinloong and Ma, Junzhang and Plumb, Nicholas C. and Radovic, Milan and Thomale, Ronny and Wilson, Stephen D. and Schnyder, Andreas P. and Shi, Ming},
doi = {10.1038/s41467-022-29828-x},
issn = {2041-1723},
journal = {Nature Communications},
month = {apr},
number = {1},
pages = {2220},
title = {{Rich nature of Van Hove singularities in Kagome superconductor CsV$_3$Sb$_5$}},
url = {https://www.nature.com/articles/s41467-022-29828-x},
volume = {13},
year = {2022}
}

@article{Khasanov2022,
author = {Khasanov, Rustem and Das, Debarchan and Gupta, Ritu and Mielke, Charles and Elender, Matthias and Yin, Qiangwei and Tu, Zhijun and Gong, Chunsheng and Lei, Hechang and Ritz, Ethan T. and Fernandes, Rafael M. and Birol, Turan and Guguchia, Zurab and Luetkens, Hubertus},
doi = {10.1103/PhysRevResearch.4.023244},
issn = {2643-1564},
journal = {Physical Review Research},
month = {jun},
number = {2},
pages = {023244},
title = {{Time-reversal symmetry broken by charge order in CsV$_3$Sb$_5$}},
url = {https://link.aps.org/doi/10.1103/PhysRevResearch.4.023244},
volume = {4},
year = {2022}
}

@article{Nie2022a,
author = {Nie, Linpeng and Sun, Kuanglv and Ma, Wanru and Song, Dianwu and Zheng, Lixuan and Liang, Zuowei and Wu, Ping and Yu, Fanghang and Li, Jian and Shan, Min and Zhao, Dan and Li, Shunjiao and Kang, Baolei and Wu, Zhimian and Zhou, Yanbing and Liu, Kai and Xiang, Ziji and Ying, Jianjun and Wang, Zhenyu and Wu, Tao and Chen, Xianhui},
doi = {10.1038/s41586-022-04493-8},
issn = {0028-0836},
journal = {Nature},
month = {apr},
number = {7904},
pages = {59--64},
title = {{Charge-density-wave-driven electronic nematicity in a kagome superconductor}},
url = {https://www.nature.com/articles/s41586-022-04493-8},
volume = {604},
year = {2022}
}

@article{Wang2023a,
author = {Wang, Jierong and Spitaler, M. and Su, Y.-S. and Zoch, K. M. and Krellner, C. and Puphal, P. and Brown, S. E. and Pustogow, A.},
doi = {10.1103/PhysRevLett.131.256501},
issn = {0031-9007},
journal = {Physical Review Letters},
month = {dec},
number = {25},
pages = {256501},
title = {{Controlled Frustration Release on the Kagome Lattice by Uniaxial-Strain Tuning}},
url = {https://link.aps.org/doi/10.1103/PhysRevLett.131.256501},
volume = {131},
year = {2023}
}

@article{Lima2023,
author = {Lima, W. P. and da Costa, D. R. and Sena, S. H. R. and Pereira, J. Milton},
doi = {10.1103/PhysRevB.108.125433},
issn = {2469-9950},
journal = {Physical Review B},
month = {sep},
number = {12},
pages = {125433},
title = {{Effects of uniaxial and shear strains on the electronic spectrum of Lieb and kagome lattices}},
url = {https://link.aps.org/doi/10.1103/PhysRevB.108.125433},
volume = {108},
year = {2023}
}

@article{Ryu2023a,
author = {Ryu, Sae Hee and Reichenbach, Garett and Jozwiak, Chris M. and Bostwick, Aaron and Richter, Peter and Seyller, Thomas and Rotenberg, Eli},
doi = {10.1016/j.elspec.2023.147357},
issn = {03682048},
journal = {Journal of Electron Spectroscopy and Related Phenomena},
month = {jul},
pages = {147357},
title = {{magnetoARPES: Angle Resolved Photoemission Spectroscopy with magnetic field control}},
url = {https://linkinghub.elsevier.com/retrieve/pii/S0368204823000749},
volume = {266},
year = {2023}
}

@article{Huang2023a,
author = {Huang, Jianwei and Yue, Ziqin and Baydin, Andrey and Zhu, Hanyu and Nojiri, Hiroyuki and Kono, Junichiro and He, Yu and Yi, Ming},
doi = {10.1063/5.0157031},
issn = {0034-6748},
journal = {Review of Scientific Instruments},
month = {sep},
number = {9},
pages = {093902},
title = {{Angle-resolved photoemission spectroscopy with an in situ tunable magnetic field}},
url = {https://pubs.aip.org/rsi/article/94/9/093902/2909975/Angle-resolved-photoemission-spectroscopy-with-an},
volume = {94},
year = {2023}
}

@article{Guo2024a,
author = {Guo, Chunyu and Wagner, Glenn and Putzke, Carsten and Chen, Dong and Wang, Kaize and Zhang, Ling and Gutierrez-Amigo, Martin and Errea, Ion and Vergniory, Maia G. and Felser, Claudia and Fischer, Mark H. and Neupert, Titus and Moll, Philip J. W.},
doi = {10.1038/s41567-023-02374-z},
issn = {1745-2473},
journal = {Nature Physics},
month = {apr},
number = {4},
pages = {579--584},
title = {{Correlated order at the tipping point in the kagome metal CsV$_3$Sb$_5$}},
url = {https://www.nature.com/articles/s41567-023-02374-z},
volume = {20},
year = {2024}
}

@article{Wei2024,
author = {Wei, Xinjian and Tian, Congkuan and Cui, Hang and Zhai, Yuxin and Li, Yongkai and Liu, Shaobo and Song, Yuanjun and Feng, Ya and Huang, Miaoling and Wang, Zhiwei and Liu, Yi and Xiong, Qihua and Yao, Yugui and Xie, X. C. and Chen, Jian-Hao},
doi = {10.1038/s41467-024-49248-3},
issn = {2041-1723},
journal = {Nature Communications},
month = {jun},
number = {1},
pages = {5038},
title = {{Three-dimensional hidden phase probed by in-plane magnetotransport in kagome metal CsV$_3$Sb$_5$ thin flakes}},
url = {https://www.nature.com/articles/s41467-024-49248-3},
volume = {15},
year = {2024}
}

@article{Liu2024,
author = {Liu, Zhaoyu and Shi, Yue and Jiang, Qianni and Rosenberg, Elliott W. and DeStefano, Jonathan M. and Liu, Jinjin and Hu, Chaowei and Zhao, Yuzhou and Wang, Zhiwei and Yao, Yugui and Graf, David and Dai, Pengcheng and Yang, Jihui and Xu, Xiaodong and Chu, Jiun-Haw},
doi = {10.1103/PhysRevX.14.031015},
issn = {2160-3308},
journal = {Physical Review X},
month = {jul},
number = {3},
pages = {031015},
title = {{Absence of $E_{2g}$ Nematic Instability and Dominant $A_{1g}$ Response in the Kagome Metal CsV$_3$Sb$_5$}},
url = {https://link.aps.org/doi/10.1103/PhysRevX.14.031015},
volume = {14},
year = {2024}
}

@article{Xing2024,
author = {Xing, Yuqing and Bae, Seokjin and Ritz, Ethan and Yang, Fan and Birol, Turan and {Capa Salinas}, Andrea N. and Ortiz, Brenden R. and Wilson, Stephen D. and Wang, Ziqiang and Fernandes, Rafael M. and Madhavan, Vidya},
doi = {10.1038/s41586-024-07519-5},
issn = {0028-0836},
journal = {Nature},
month = {jul},
number = {8019},
pages = {60--66},
title = {{Optical manipulation of the charge-density-wave state in RbV$_3$Sb$_5$}},
url = {https://www.nature.com/articles/s41586-024-07519-5},
volume = {631},
year = {2024}
}

@article{Guo2024,
author = {Guo, Chunyu and Wagner, Glenn and Putzke, Carsten and Chen, Dong and Wang, Kaize and Zhang, Ling and Gutierrez-Amigo, Martin and Errea, Ion and Vergniory, Maia G. and Felser, Claudia and Fischer, Mark H. and Neupert, Titus and Moll, Philip J. W.},
doi = {10.1038/s41567-023-02374-z},
issn = {1745-2473},
journal = {Nature Physics},
month = {apr},
number = {4},
pages = {579--584},
title = {{Correlated order at the tipping point in the kagome metal CsV$_3$Sb$_5$}},
url = {https://www.nature.com/articles/s41567-023-02374-z},
volume = {20},
year = {2024}
}

@article{Asaba2024,
author = {Asaba, T. and Onishi, A. and Kageyama, Y. and Kiyosue, T. and Ohtsuka, K. and Suetsugu, S. and Kohsaka, Y. and Gaggl, T. and Kasahara, Y. and Murayama, H. and Hashimoto, K. and Tazai, R. and Kontani, H. and Ortiz, B. R. and Wilson, S. D. and Li, Q. and Wen, H. -H. and Shibauchi, T. and Matsuda, Y.},
doi = {10.1038/s41567-023-02272-4},
issn = {1745-2473},
journal = {Nature Physics},
month = {jan},
number = {1},
pages = {40--46},
title = {{Evidence for an odd-parity nematic phase above the charge-density-wave transition in a kagome metal}},
url = {https://www.nature.com/articles/s41567-023-02272-4},
volume = {20},
year = {2024}
}

@article{Tazai2024,
author = {Tazai, Rina and Yamakawa, Youichi and Kontani, Hiroshi},
doi = {10.1073/pnas.2303476121},
issn = {0027-8424},
journal = {Proceedings of the National Academy of Sciences},
month = {jan},
number = {3},
title = {{Drastic magnetic-field-induced chiral current order and emergent current-bond-field interplay in kagome metals}},
url = {https://pnas.org/doi/10.1073/pnas.2303476121},
volume = {121},
pages = {e2303476121},
year={2024}
}

@article{Wang2024,
author = {Wang, Jingyuan and Farhang, Camron and Ortiz, Brenden R. and Wilson, Stephen D. and Xia, Jing},
doi = {10.1103/PhysRevMaterials.8.014202},
issn = {2475-9953},
journal = {Physical Review Materials},
month = {jan},
number = {1},
pages = {014202},
title = {{Resolving the discrepancy between MOKE measurements at 1550-nm wavelength on kagome metal CsV$_3$Sb$_5$}},
url = {https://link.aps.org/doi/10.1103/PhysRevMaterials.8.014202},
volume = {8},
year = {2024}
}

@article{Saykin2023,
author = {Saykin, David R. and Farhang, Camron and Kountz, Erik D. and Chen, Dong and Ortiz, Brenden R. and Shekhar, Chandra and Felser, Claudia and Wilson, Stephen D. and Thomale, Ronny and Xia, Jing and Kapitulnik, Aharon},
doi = {10.1103/PhysRevLett.131.016901},
issn = {0031-9007},
journal = {Physical Review Letters},
month = {jul},
number = {1},
pages = {016901},
title = {{High Resolution Polar Kerr Effect Studies of CsV$_3$Sb$_5$: Tests for Time-Reversal Symmetry Breaking below the Charge-Order Transition}},
url = {https://link.aps.org/doi/10.1103/PhysRevLett.131.016901},
volume = {131},
year = {2023}
}

@article{Farhang2023,
author = {Farhang, Camron and Wang, Jingyuan and Ortiz, Brenden R. and Wilson, Stephen D. and Xia, Jing},
doi = {10.1038/s41467-023-41080-5},
issn = {2041-1723},
journal = {Nature Communications},
month = {sep},
number = {1},
pages = {5326},
title = {{Unconventional specular optical rotation in the charge ordered state of Kagome metal CsV$_3$Sb$_5$}},
url = {https://www.nature.com/articles/s41467-023-41080-5},
volume = {14},
year = {2023}
}

@article{Xiang2021,
author = {Xiang, Ying and Li, Qing and Li, Yongkai and Xie, Wei and Yang, Huan and Wang, Zhiwei and Yao, Yugui and Wen, Hai-Hu},
doi = {10.1038/s41467-021-27084-z},
issn = {2041-1723},
journal = {Nature Communications},
month = {nov},
number = {1},
pages = {6727},
title = {{Twofold symmetry of c-axis resistivity in topological kagome superconductor CsV$_3$Sb$_5$ with in-plane rotating magnetic field}},
url = {https://www.nature.com/articles/s41467-021-27084-z},
volume = {12},
year = {2021}
}

@article{Jin2024,
author = {Jin, Feng and Ren, Wei and Tan, Mingshu and Xie, Mingtai and Lu, Bingru and Zhang, Zheng and Ji, Jianting and Zhang, Qingming},
doi = {10.1103/PhysRevLett.132.066501},
issn = {0031-9007},
journal = {Physical Review Letters},
month = {feb},
number = {6},
pages = {066501},
title = {{$\pi$ Phase Interlayer Shift and Stacking Fault in the Kagome Superconductor CsV$_3$Sb$_5$}},
url = {https://link.aps.org/doi/10.1103/PhysRevLett.132.066501},
volume = {132},
year = {2024}
}

@article{Guo2009a,
author = {Guo, H.-M. and Franz, M.},
doi = {10.1103/PhysRevB.80.113102},
issn = {1098-0121},
journal = {Physical Review B},
month = {sep},
number = {11},
pages = {113102},
title = {{Topological insulator on the kagome lattice}},
url = {https://link.aps.org/doi/10.1103/PhysRevB.80.113102},
volume = {80},
year = {2009}
}

@article{Christensen2021,
  title = {{Theory of the charge density wave in $A{\mathrm{V}}_{3}{\mathrm{Sb}}_{5}$ kagome metals}},
  author = {Christensen, Morten H. and Birol, Turan and Andersen, Brian M. and Fernandes, Rafael M.},
  journal = {Phys. Rev. B},
  volume = {104},
  issue = {21},
  pages = {214513},
  numpages = {15},
  year = {2021},
  month = {Dec},
  publisher = {American Physical Society},
  doi = {10.1103/PhysRevB.104.214513},
  url = {https://link.aps.org/doi/10.1103/PhysRevB.104.214513}
}

@article{Park2021,
  title = {{Electronic instabilities of kagome metals: Saddle points and Landau theory}},
  author = {Park, Takamori and Ye, Mengxing and Balents, Leon},
  journal = {Phys. Rev. B},
  volume = {104},
  issue = {3},
  pages = {035142},
  numpages = {20},
  year = {2021},
  month = {Jul},
  publisher = {American Physical Society},
  doi = {10.1103/PhysRevB.104.035142},
  url = {https://link.aps.org/doi/10.1103/PhysRevB.104.035142}
}

@article{Lin2021,
  title = {{Complex charge density waves at Van Hove singularity on hexagonal lattices: Haldane-model phase diagram and potential realization in the kagome metals $A{V}_{3}{\mathrm{Sb}}_{5}$ ($A$=K, Rb, Cs)}},
  author = {Lin, Yu-Ping and Nandkishore, Rahul M.},
  journal = {Phys. Rev. B},
  volume = {104},
  issue = {4},
  pages = {045122},
  numpages = {13},
  year = {2021},
  month = {Jul},
  publisher = {American Physical Society},
  doi = {10.1103/PhysRevB.104.045122},
  url = {https://link.aps.org/doi/10.1103/PhysRevB.104.045122}
}

@article{Christensen2022,
  title = {{Loop currents in $A{\mathrm{V}}_{3}{\mathrm{Sb}}_{5}$ kagome metals: Multipolar and toroidal magnetic orders}},
  author = {Christensen, Morten H. and Birol, Turan and Andersen, Brian M. and Fernandes, Rafael M.},
  journal = {Phys. Rev. B},
  volume = {106},
  issue = {14},
  pages = {144504},
  numpages = {18},
  year = {2022},
  month = {Oct},
  publisher = {American Physical Society},
  doi = {10.1103/PhysRevB.106.144504},
  url = {https://link.aps.org/doi/10.1103/PhysRevB.106.144504}
}

@article{Denner2021,
  title = {{Analysis of Charge Order in the Kagome Metal $A{\mathrm{V}}_{3}{\mathrm{Sb}}_{5}$ ($A=\mathrm{K},\mathrm{Rb},\mathrm{Cs}$)}},
  author = {Denner, M. Michael and Thomale, Ronny and Neupert, Titus},
  journal = {Phys. Rev. Lett.},
  volume = {127},
  issue = {21},
  pages = {217601},
  numpages = {6},
  year = {2021},
  month = {Nov},
  publisher = {American Physical Society},
  doi = {10.1103/PhysRevLett.127.217601},
  url = {https://link.aps.org/doi/10.1103/PhysRevLett.127.217601}
}

@article{Fischer2023,
  title = {Phenomenology of bond and flux orders in kagome metals},
  author = {Wagner, Glenn and Guo, Chunyu and Moll, Philip J. W. and Neupert, Titus and Fischer, Mark H.},
  journal = {Phys. Rev. B},
  volume = {108},
  issue = {12},
  pages = {125136},
  numpages = {17},
  year = {2023},
  month = {Sep},
  publisher = {American Physical Society},
  doi = {10.1103/PhysRevB.108.125136},
  url = {https://link.aps.org/doi/10.1103/PhysRevB.108.125136}
}

@article{Ritz2023_SC,
  title = {{Superconductivity from orbital-selective electron-phonon coupling in $A{\mathrm{V}}_{3}{\mathrm{Sb}}_{5}$}},
  author = {Ritz, Ethan T. and R\o{}ising, Henrik S. and Christensen, Morten H. and Birol, Turan and Andersen, Brian M. and Fernandes, Rafael M.},
  journal = {Phys. Rev. B},
  volume = {108},
  issue = {10},
  pages = {L100510},
  numpages = {7},
  year = {2023},
  month = {Sep},
  publisher = {American Physical Society},
  doi = {10.1103/PhysRevB.108.L100510},
  url = {https://link.aps.org/doi/10.1103/PhysRevB.108.L100510}
}

\section{Methods}
\subsection{Sample growth and characterization}
Single crystals of \cvs~were grown with self-flux method described in Ref.~\cite{Qian2021} and~\cite{Liu2024}. The crystal sizes are 3 $\times$ 3 mm with typical residual resistivity ratio (RRR) up to $\sim$120. The resistivity measurements were performed in DynaCool (Quantum Design, Inc.) with Stanford Research SR860 Lock-in amplifiers and voltage preamplifiers.

\subsection{MagnetoARPES measurements}
ARPES experiments were performed using a laboratory system equipped with a helium lamp light source that is 75\% linear vertical and 25\% linear horizontal polarization and a Scienta DA30 electron analyzer. The angular resolution was set to 0.3$^\circ$. The total energy resolution was set to 13 meV. All the samples were cleaved \textit{in situ} at around 30 K and all the measurements were conducted in an ultra-high vacuum (UHV) with a base pressure lower than 1 $\times$ 10$^{-10}$ Torr. Results were reproduced across multiple samples. 

An \textit{in situ} magnetic field was applied using a magnetic coil mounted around the sample on the sample stage (Fig.~\ref{fig:Fig1}e and f)~\cite{Huang2023a}. Samples were placed at the center of the magnetic coil flush with the top surface of the coil so that the magnetic field was oriented along the out-of-plane direction at the sample position. The magnitude of the field was controlled by adjusting the electric current flowing through the magnetic coil, which was characterized outside the UHV prior to the experiments~\cite{Huang2023a}. The magnetic field direction can be reversed by reversing the direction of the electrical current. In this experimental setup, the primary extrinsic effects of the magnetic field on ARPES measurements are constant energy contour rotation, photoelectron emission angle contraction, and momentum broadening~\cite{Huang2023a}. In this study, with the small applied magnetic field, only a rigid constant energy contour rotation was observed, which could be corrected post-measurements. All the ARPES measurements were performed using the DA30 deflector mode unless otherwise noted.

\subsection{DFT calculations}
The density functional theory-based first-principles calculations were performed with the Vienna $ab$-$initio$ Simulation Package \cite{Kresse1996b}. The generalized gradient approximation parameterized by Perdew, Burke, and Ernzerhof (PBE) \cite{Perdew1997} was employed for the exchange-correlation interaction between electrons throughout. The plane-wave basis set cutoff energy was 300 eV. The pristine phase of CsV$_3$Sb$_5$ was used to calculate the electronic structures. Notice that the crystal structure was obtained from Ref.~\cite{Tan2021}. Spin-orbit coupling was not involved in structural relaxation but was considered in band structure calculation. A $k$-mesh of 12$\times$12$\times$6 was utilized to sample the Brillouin zone.


\section{Simulation of magneto-dichroism of ARPES spectra}
We observed magneto-dichroism in the ARPES spectra between the two adjacent K regions when applying opposite magnetic fields along the $c$-axis (Fig. 2). We have shown in the main text that the magneto-dichroism effect arises from the selective broadening of the bands at different K regions (Fig. 2, 3). The effect can be qualitatively simulated with the single-particle spectral function $A(k,\omega)$ by manually assigning different imaginary components to the self-energy for bands around the two K regions (Extended Data Fig. 3). $A(k,\omega)$ is described by:
\begin{equation}
A(k,\omega) = -\frac{1}{\pi} \frac{\Sigma_{2}(k,\omega)}{[\omega-\epsilon_k-\Sigma_{1}(k,\omega)]^2+\Sigma_{2}(k,\omega)^2}
\end{equation}

$\epsilon_k$ is taken from the DFT calculations (dashed line in Extended Data Fig. 3). $\Sigma_{2}(k,\omega)$ is obtained by fitting the linewidth of the corresponding experimental data. A momentum- and energy- independent $\Sigma_{2}(k,\omega)$ is adopted for the simulation. The entire spectrum is then multiplied by a Fermi distribution at 25 K. With applying a positive magnetic field, $\Sigma_{2}(k,\omega)$ values of 0.21 eV and 0.15 eV were individually used for $A(k,\omega)$, corresponding to the simulated APRES spectra around different K regions (left and right part of Extended Data Fig. 3a and d). When the magnetic field is reversed, the opposite $\Sigma_{2}(k,\omega)$ values are used. The simulated spectra exhibit a similar magneto-dichroism effect as observed in the experimental data (Extended Data Fig. 1).


\subsection{\edt{One-step model ARPES calculations}}
\edt{To understand the matrix element effects in ARPES experiments under external magnetic fields, we carried out ab-initio based one-step model ARPES calculations using the spin-polarized relativistic Korringa-Kohn-Rostoker (SPRKKR) package~\cite{Ebert2011Calculating} under the fully relativistic four component Dirac formalism and the atomic-sphere approximation. The chosen exchange-correlation functional is based on the local spin density approximation by Vosko, Wilk, and Nusair~\cite{Vosko1980Accurate}. An angular momentum cutoff of l=3 was adopted for solving the KKR equations. During the self-consistent field (SCF) calculations, the Brillouin zone integration was carried out on an equidistant 29×29×14 k-mesh and the energy integral was evaluated with 32 points along the Gaussian-Legendre quadrature path. For each external magnetic field, the SCF calculations were converged separately. The Lloyd’s formula was employed to determine the Fermi level. The one-step ARPES calculations considered a semi-infinite surface model terminated by the Sb atoms with the next layer followed by the V-Sb slab, rather than Cs atoms. The surface barrier takes the z-dependent Rundgren-Malmström type~\cite{Rundgren1977Transmission}. Experimental light incidence angles and the photon energy of 21.2 eV were adopted in the calculations. To simulate the mixed polarization of the helium lamp, the final calculated one-step ARPES spectra were obtained through a linear superposition of 25\% p-polarization and 75\% s-polarization results. The life-time effects of the initial and final states were simulated by an imaginary potential of 0.05 eV and 1.5 eV, respectively. The external magnetic field theoretically enters all three parts of the photoemission matrix elements—initial states, the relativistic dipole operator, and the spin-polarized time-reversed low-energy electron diffraction final states, the treatment of which has been described previously for ferromagnetic materials~\cite{Braun1996Theory}. The classical photoelectron deflection in the vacuum due to the Lorentz force under magnetic fields was not considered in the one-step model ARPES calculations.}

\edt{The results of the calculations are summarized in Extended Data Fig. 5. First, we show the zero-field constant energy contour (CEC) 260 meV below the Fermi level (Extended Data Fig. 5a). Notice that the Fermi level in the calculations is approximately 90 meV higher than the one in experiments. Therefore, the theoretical CEC in Extended Data Fig. 5a presents reasonable resemblance to the experimental $E$ – \ef~= -0.35 eV CEC of Fig. 2(a) taken slightly below the Dirac cone. Without adjusting the theoretical Fermi level, we extract the momentum slice Cut 1 at $k_x = 0.5$ \AA$^{-1}$ and compare its band dispersions and MDCs under opposite directions of large external fields around 117.5 T. The results are summarized in Extended Data Fig. 5b-d. The horizontal dashed lines near the MDC (taken at $E$ – \ef~= -0.2 eV) inner peaks clearly indicate a left-right symmetric feature at 0 T, but field-odd asymmetric behavior under the applied large magnetic fields. While this may seem consistent with the experimental data in Fig. 2e-g, the fields required to observe this asymmetry in the simulations are five orders of magnitude larger than the experimental ones. The difference in the theoretical calculations here is that the TRS-breaking comes from the large magnetic fields and the induced moments on V atoms, treated within the DFT. To gain further insights on the asymmetric MDC peaks in Extended Data Fig. 5c-d, we present the spin-resolved MDCs under different magnetic fields in Extended Data Fig. 5e-g with the projection axis chosen along z. Since if only considering the out-of-plane magnetic field which breaks the time reversal symmetry and all vertical mirrors, the magnetic mirror $M_xT$ ($\Gamma$-K containing) preserves, protecting the band degeneracy between $E(k_x,k_y)$ and $E(k_x,-k_y)$. Therefore, we infer that spin-dependent multiple scattering and relativistic effects could contribute to the asymmetric MDC line-shapes with respect to the $\Gamma$-M plane. However, this only becomes discernible in the calculations with magnetic fields nearly five orders of magnitude larger than the experimental ones. As a comparison, we show in Extended Data Fig. 5h the Cut 1 spin-integrated MDCs taken at $E$ – \ef~= -0.2 eV under 0, +2.35, and -2.35 T fields. The calculated results clearly overlap with each other without any signatures of the magneto-dichroism observed in ARPES experiments.}


\subsection{Model for the fingerprints of piezomagnetism in ARPES}
We discuss a phenomenological model to shed light on the manifestations of piezomagnetism in our magneto-ARPES data. Specifically, we consider the congruent CDW flux phase proposed in Ref. \cite{Xing2024}, which has the non-zero piezomagnetic tensor component $\Lambda_{xyz}$, defined as $\varepsilon_{ij} = \Lambda_{ijk}B_k$, where $\varepsilon_{ij}$ is the strain tensor and $B_k$ is the magnetic field. Thus, $\Lambda_{xyz}$ implies that a non-zero shear strain $\varepsilon_{xy}$ is induced by an out-of-plane field $B_z$. Importantly, the congruent CDW flux phase also displays a non-zero distortion $\varepsilon_{xx} - \varepsilon_{xy}$ in the absence of a magnetic field. Our goal is to model the impact of both effects on the V and Sb Fermi surfaces. 

We start with the large vanadium kagome Fermi surface. For simplicity, we model it in terms of a simple tight-binding model on the kagome lattice. Denoting the three kagome sublattices as A, B, and C, and following the coordinate system of Ref.~\cite{Guo2009a}, the nearest-neighbor tight-binding model is given by:
\begin{equation}
\mathcal{H}_0 = \sum_{\bm{k},\sigma} \Psi^\dagger_{\bm{k}\sigma}H_0(\bm{k})\Psi_{\bm{k}\sigma}
\end{equation}
where $\Psi_{\bm{k}\sigma} = (\psi_{A,\bm{k}\sigma}, \psi_{B,\bm{k}\sigma}, \psi_{C,\bm{k}\sigma})^T$ and:
\begin{eqnarray}
H_0(\bm{k}) = -2t
\left( \begin{array}{ccc}
0 & \text{cos}~k_1 & \text{cos}~k_2 \\ 
\text{cos}~k_1 & 0 & \text{cos}~k_3 \\ 
\text{cos}~k_2 & \text{cos}~k_3 & 0 \\
\end{array}\right)
-\mu
\left( \begin{array}{ccc}
1 & 0 & 0 \\ 
0 & 1 & 0 \\ 
0 & 0 & 1 \\
\end{array}\right)
\end{eqnarray}

 Here, $\mu$ denotes the chemical potential, $t$ is the hopping parameter and $k_1 = k_x$, $k_2 = \dfrac{k_x}{2} + \dfrac{\sqrt{3}k_y}{2}$, $k_3 = -\dfrac{k_x}{2} + \dfrac{\sqrt{3}k_y}{2}$. The band structure has a saddle point at $\mu$ = 0, resulting in a perfectly hexagonal Fermi surface. 

In the congruent CDW flux phase proposed in Ref.~\cite{Xing2024}, two additional terms emerge:
\begin{equation}
\mathcal{H} = \mathcal{H}_0 + \mathcal{H}_{1}
\end{equation}

with:
\begin{eqnarray} \label{eq:largeFS}
H_{1}(\bm{k}) = \frac{\Phi_1}{\sqrt{3}}
\left( \begin{array}{ccc}
1 & 0 & 0 \\ 
0 & 1 & 0 \\ 
0 & 0 &-2 \\
\end{array}\right)
+ \Phi_2B_z
\left( \begin{array}{ccc}
1 & 0 & 0 \\ 
0 & -1 & 0 \\ 
0 & 0 & 0\\
\end{array}\right)
\end{eqnarray}

Here, $\Phi_1$ and $\Phi_2$ are parameters that transform as the $d_{x^2-y^2}$ and $d_{xy}$ form factors, respectively, and $B_z$ is the out-of-plane magnetic field. Note that this parametrization gives one of the domains only. While both terms individually break $C_6$ symmetry, their combination also breaks all in-plane twofold rotation axes (i.e., all vertical mirrors). We can diagonalize $H_0(\bm{k}) + H_{1}(\bm{k})$ to obtain the distorted Fermi surface under a magnetic field. We note that because the congruent CDW flux phase breaks translational symmetry, it also folds the band structure. We do not include these effects since our goal is to focus on the piezomagnetic effect.

The fact that the ARPES data without a magnetic field is $C_6$ symmetric indicates that either $\Phi_1 = 0$ or $\Phi_1 \ll t$, as shown in the Extended Data Fig. 6. For the calculations incorporating a non-zero magnetic field, we adopted $\Phi_1 = 0.01t$. 

To visualize the behavior of the Fermi surface in the presence of a magnetic field with this model, we set $\mu = 0.16 t$ (in order to be slightly away from perfect nesting) and calculate the Fermi surface with the parameters ($t$, $\Phi_1$, $\Phi_2B_z$) = (1, 0.01, 0.1) to represent a positive magnetic field and ($t$, $\Phi_1$, $\Phi_2B_z$) = (1, 0.01, -0.1) to represent a negative magnetic field (Extended Data Fig. 7). The Fermi surfaces obtained in the presence of a magnetic field for both situations preserve inversion symmetry but break the mirror symmetry with respect to the horizontal $\Gamma$--M plane (Extended Data Fig. 7b and c). This is consistent with the $C_6$ symmetry-breaking behavior observed in the experimental data, which however is manifested in the spectral weight distribution of the Fermi surface.

For the case of the small Sb Fermi pocket, it suffices to model the electronic dispersion in terms of a parabolic dispersion with two broken-symmetry terms analogous to $\Phi_1$ and $\Phi_2$ above in Eq. (\ref{eq:largeFS}):

\begin{equation}{\label{eq:smallFS}}
E(k,\theta) = \left( \frac{k^2}{2m} - \tilde{\mu} \right) + \tilde{\Phi}_1 k^2 \cos (2\theta) + \tilde{\Phi}_2 B_z k^2 \sin (2\theta)
\end{equation}
where $\theta$ is the angle with respect to the $k_x$ axis, which corresponds to a $\Gamma - M$ direction. We use this equation to fit the angle dependence of the Fermi surface diameter extracted from the ARPES data. To simplify the notation, we rewrite the equation above as follows:

\begin{equation}{\label{equ:EqFit}}
E(k,\theta) = (A_1 k^2 - \mu) + A_2 k^2 \cos (2\theta) + A_3 k^2 \sin (2\theta)
\end{equation}

To accurately capture the Fermi surface distortion under a magnetic field, we first extract the Fermi surface diameter as a function of the Fermi surface angle by fitting the Fermi surface mappings obtained with and without a magnetic field (Extended Data Fig. 8a). We then normalize the Fermi surface diameter under a magnetic field by dividing it by the diameter measured without a magnetic field, thus excluding potential extrinsic effects. The normalized Fermi surface diameter as a function of the Fermi surface angle exhibits a more pronounced two-fold rotational symmetry (Fig.~\ref{fig:Fig4}j). We then fit the Fermi momentum using Eq.~\ref{equ:EqFit} (Extended Data Fig. 8b), yielding fitting parameters $A_2/A_1 \sim 0.050$ and $A_3/A_1 \sim 0.011$. The fitting results reveal a consistent angular deviation of the longitudinal axis of the elliptical Fermi surface from the BZ high-symmetry direction, in agreement with the fitting approach used in the main text.

\section{Data Availability}
All data needed to evaluate the conclusions are present in the paper and extended data. Additional data are available from the corresponding authors upon reasonable request.

\section{Code Availability}
The band structure calculations used in this study are available from the corresponding authors upon reasonable request.

\section{Acknowledgments}
The ARPES work at Rice University was supported by the Gordon and Betty Moore Foundation's EPiQS Initiative through grant No. GBMF9470 (M.Y.), the Robert A. Welch Foundation Grant No. C-2175 (M.Y.), and the U.S. Department of Energy, Basic Energy Sciences Grant Nos. DF-SC0021421 and DE-SC0026179 (M.Y.).
The single-crystal synthesis work at UW was supported by the Air Force Office of Scientific Research (AFOSR) under Award No. FA2386-21-1-4060 (J.H.C.) and the David Lucile Packard Foundation (J.H.C.). R.M.F. was supported by the Air Force Office of Scientific Research under Award No. FA9550-21-1-0423. Z.W. is supported by the Department of Energy, Basic Energy Sciences Grant No. DE-FG02-99ER45747. B.Y. acknowledges the financial support by the Israel Science Foundation (2974/23) and National Science Foundation through the Penn State Materials Research Science and Engineering Center (MRSEC) DMR 2011839. Single crystal growth efforts at Rice is supported by  the U.S. DOE, BES DE-SC0026179 (P.D.). Part of the materials characterization efforts at Rice is supported by the Robert A. Welch Foundation Grant No. C-1839 (P.D.). J.K. acknowledges support from the Robert A. Welch Foundation (through Grant No. C-1509) and the Gordon and Betty Moore Foundation (through Grant No. 11520). J.M. and A.P. thank the project Quantum materials for applications in sustainable technologies (QM4ST), funded as Project No. CZ.02.01.01$/$00$/$22$\_$008$/$0004572 by Programme Johannes Amos Comenius, call Excellent Research. Y.H. is supported by Air Force Office of Scientific Research under Award No. FA9550-24-1-0048.

\section{Author contributions}
M.Y. and J.Huang proposed and designed the project. Z.L. grew the \cvs~single crystals with the help of J.M.D. under the guidance of J.H.C.. J.Huang, Z.R., J.Hyun, T.H. and Z.Y. carried out the ARPES measurements under the guidance of M.Y. Y.H. and J.K.. J.Huang performed the magneto-ARPES data analysis. Y.X., Z.L., and J.M.D. grew single crystals for control measurements under the guidance of P.D. and J.H.C.. H.T., B.Y. and Z.W. conducted the first-principles calculations. Y.Z. carried out the ab-initio based one-step model ARPES calculations with advice from A.P. and J.M.. Y.Z. analyzed the one-step calculated data. R.M.F. performed the phenomenological theoretical analysis. J.Huang and M.Y. wrote the paper with input from all co-authors.

\section{Competing interests}
The authors declare that they have no competing interests.

\newpage

\begin{figure}
\includegraphics[width=1\textwidth]{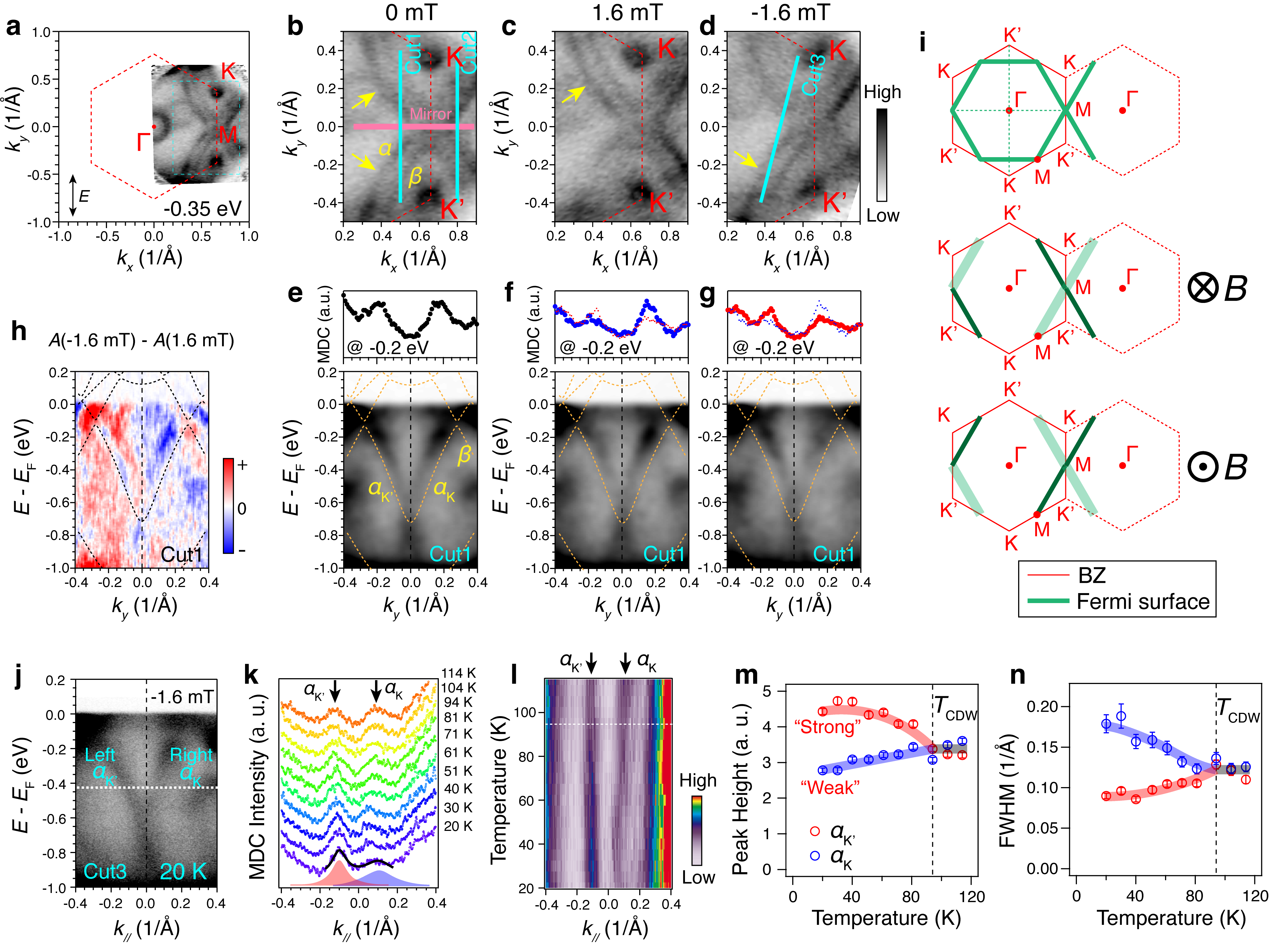}
\caption{\label{fig:Fig2}{\bf Field response of the electronic states originating from the V $d$-orbitals near K/K'.} (a) Constant energy contour of \cvs~at -0.35~eV measured without a magnetic field using DA30 mode. The double-headed arrow indicates light polarization. (b) Zoom-in view within the cyan rectangle in (a). The yellow arrows point to the sharp electronic sheets around both K and K'. (c) Same as (b) but in the presence of a magnetic field of +1.6~mT. The yellow arrow points to the electronic sheets that remain sharp around K. (d) Same as (b) but in the presence of a magnetic field of -1.6~mT. The yellow arrow points to the electronic sheets that remain sharp around K'. (e)-(g) $E-k$ spectral images and the related momentum distribution curves (MDCs) at -0.2~eV along Cut1 indicated in (b), corresponding to the measurements under a magnetic field of 0~mT, +1.6~mT and -1.6~mT, respectively. The yellow dashed lines overlaid are the corresponding DFT band dispersions. (h) The magneto-dichroic spectral image obtained by subtracting (f) from (g). The black dashed lines overlaid are the DFT band dispersions. (i) Schematics showing the VHS Fermi surfaces without and with a magnetic field. The thicker and fainter lines indicate broadened branches, while thinner and more intense lines represent sharper branches. }
\end{figure}
\addtocounter{figure}{-1}
\begin{figure}
\caption{(continued) 
(j) Spectral images taken along Cut3 marked in (d) measured at -1.6~mT and 20~K. (k) MDCs taken at -0.42~eV as marked in (j) measured at various temperatures. The fitted peaks for $\alpha_{K'}$ and $\alpha_{K}$ bands measured at 20~K are shown as shaded peaks. (l) Spectral image obtained by combining the MDC stacks in (k). (m) Fitted peak height and (n) peak width of $\alpha_{K'}$ and $\alpha_{K}$ as a function of temperature.
}
\end{figure}

\begin{figure}
\includegraphics[width=1\textwidth]{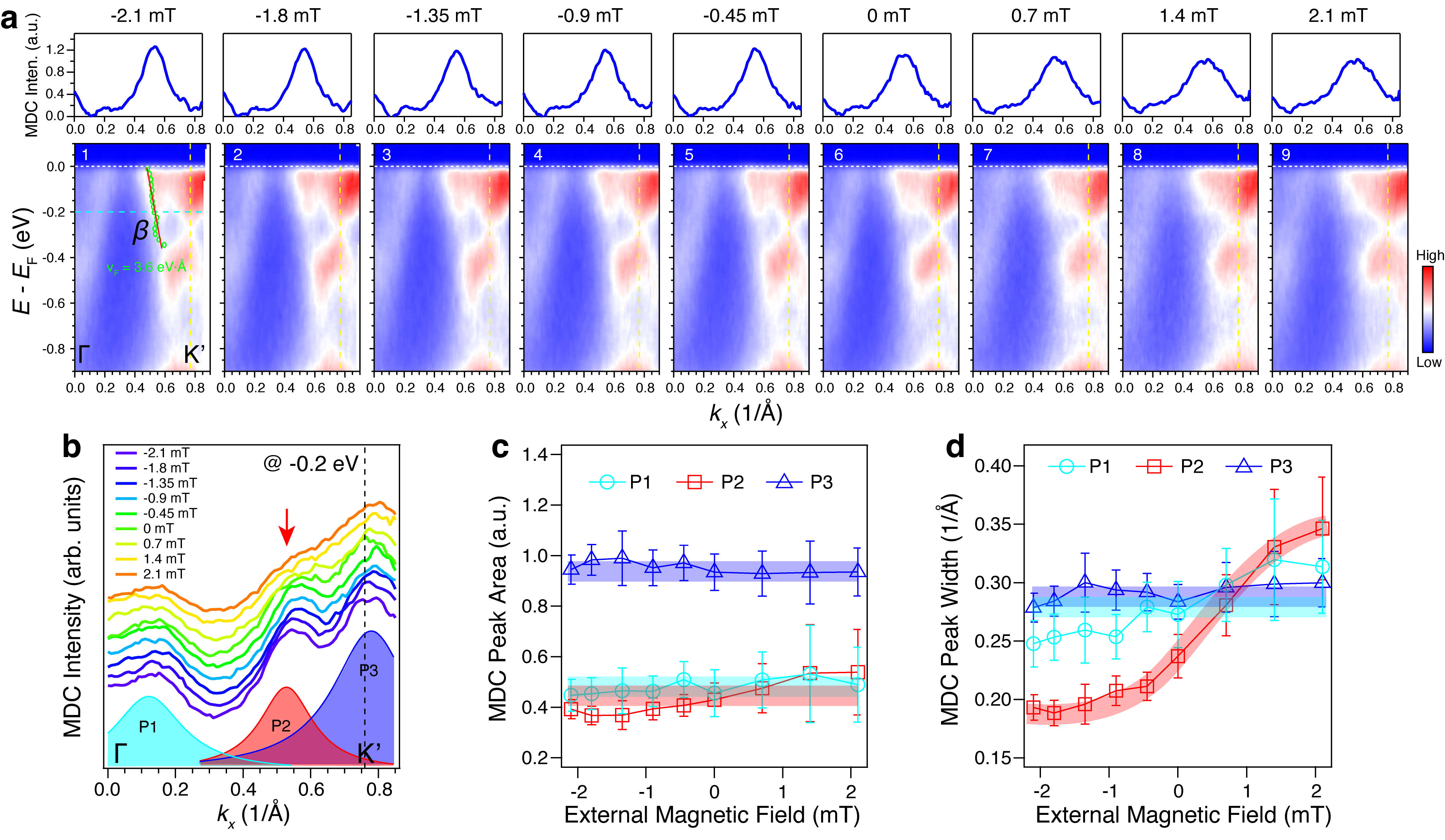} 
\caption{\label{fig:Fig3}{\bf Magnetic field dependence of the electronic spectrum near K.} (a) Spectral images along the $\Gamma$ - K' direction at different external magnetic fields. The top row shows the corresponding P2 peaks indicated in (b) after subtracting the fitted background and other peaks from the momentum distribution curves (MDCs). (b) Corresponding MDCs at -0.2 eV of the spectral images in (a) at different magnetic fields. The red arrow points to the peak (P2) from the $\beta$ band which responds most prominently to the magnetic field. The $\alpha$ band is strongly suppressed in this measurement geometry due to matrix element effects. The MDCs are fitted by three Lorentzian peaks (P1, P2 and P3) and a fixed constant background. (c) The fitted MDC peak areas as a function of magnetic field for the three corresponding peaks in (b). (d) The fitted MDC peak widths as a function of magnetic field for the three corresponding peaks in (b).
}
\end{figure}

\begin{figure}
\includegraphics[width=1\textwidth]{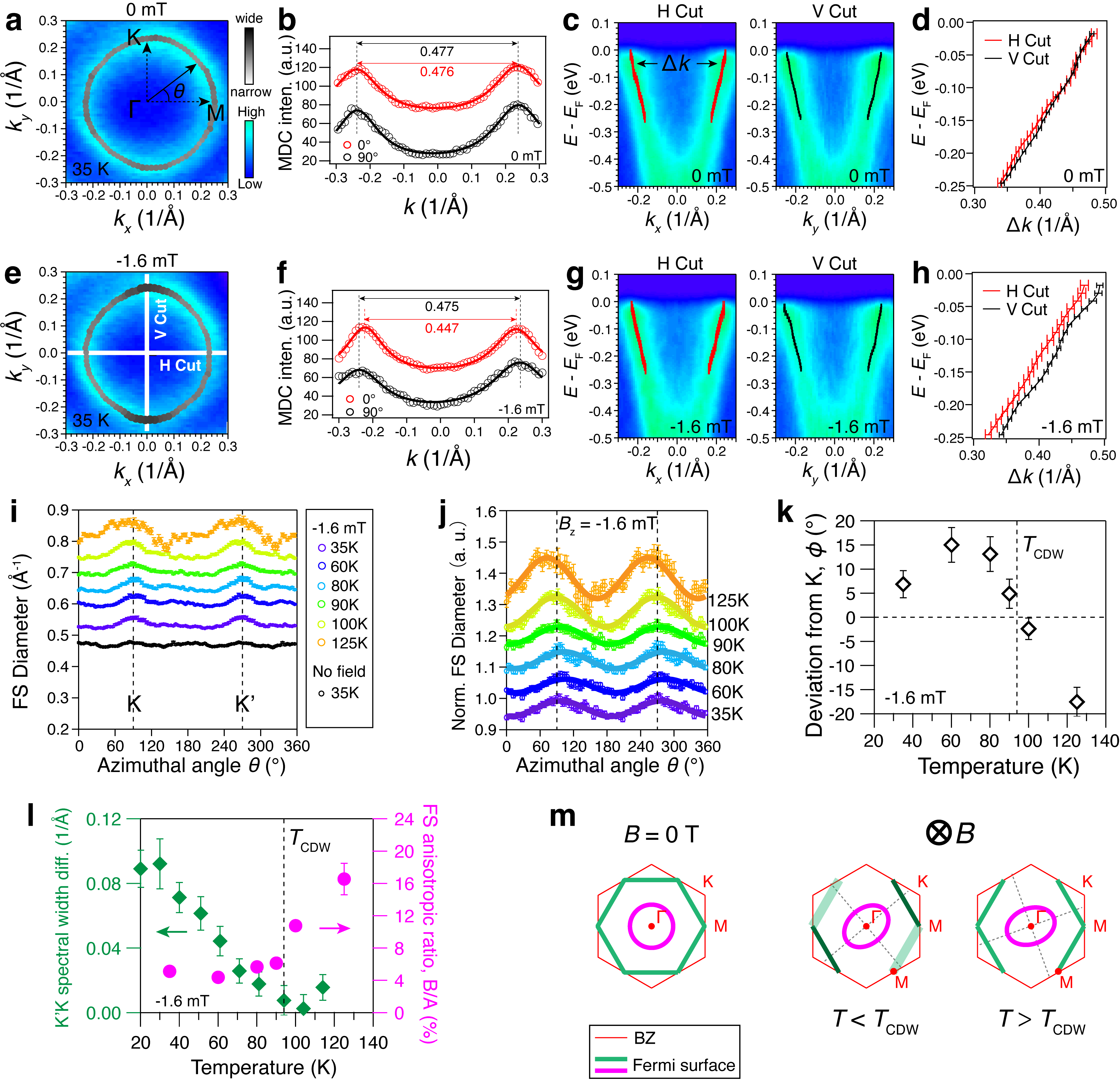}
\caption{\label{fig:Fig4} {\bf Field response of the electronic states originating from the Sb $p$-orbitals near \g.} (a) Measured Fermi surface around $\Gamma$ in the absence of a magnetic field. The azimuthal angle $\theta$ is defined with respect to the horizontal \g-M direction. (b) MDC across the Fermi pocket along the horizontal ($\theta=0^{\circ}$) and vertical ($\theta=90^{\circ}$) directions, fitted by a two-Lorentzian function to extract the diameter measured along the two directions. (c) Spectral images measured along the two orthogonal cuts, overlaid with the peak positions of the MDC fitting for the dispersions. (d) The extracted momentum separation of the fitted dispersions in (c) as a function of energy. (e)-(h) Same as (a)-(d) but for measurements in a magnetic field of -1.6 mT. (i) Fermi pocket diameter as a function of the azimuthal angle obtained by fitting the radial MDCs without and with a magnetic field at different temperatures.
}
\end{figure}
\addtocounter{figure}{-1}
\begin{figure}
\caption{(continued) 
 The curves are offset for clarity. (j) Fermi pocket diameter for -1.6 mT at different temperatures normalized by its zero-field value measured at 35 K. All curves are fitted with the function $A + B \cos(2(\theta+\phi))$. (k) The deviation of the long axis of the elliptical Fermi pocket away from the K point as a function of temperature at -1.6 mT. (l) Temperature dependence of the ellipticity of the Fermi pocket (pink) plotted together with the temperature evolution of the VHS spectral difference, all measured at -1.6 mT. (m) Summary schematics of the temperature evolution of the field response of the electronic structure of \cvs. At zero magnetic field for any temperature, the \g~pocket is circular and the large Fermi surface associated with the VHS respect the C$_6$ symmetry. In the presence of a magnetic field, for $T$ < \tcdw, the large Fermi surface break the C$_6$ rotational symmetry while the \g~pocket becomes elongated and rotates away from \g - K. Just above \tcdw, the large Fermi surface restore their C$_6$ symmetry but the \g~pocket retains its ellipticity.
}
\end{figure}

\newpage
\setcounter{figure}{0}
\renewcommand{\figurename}{Extended Data Fig.}

\begin{figure}
\includegraphics[width=0.95\textwidth]{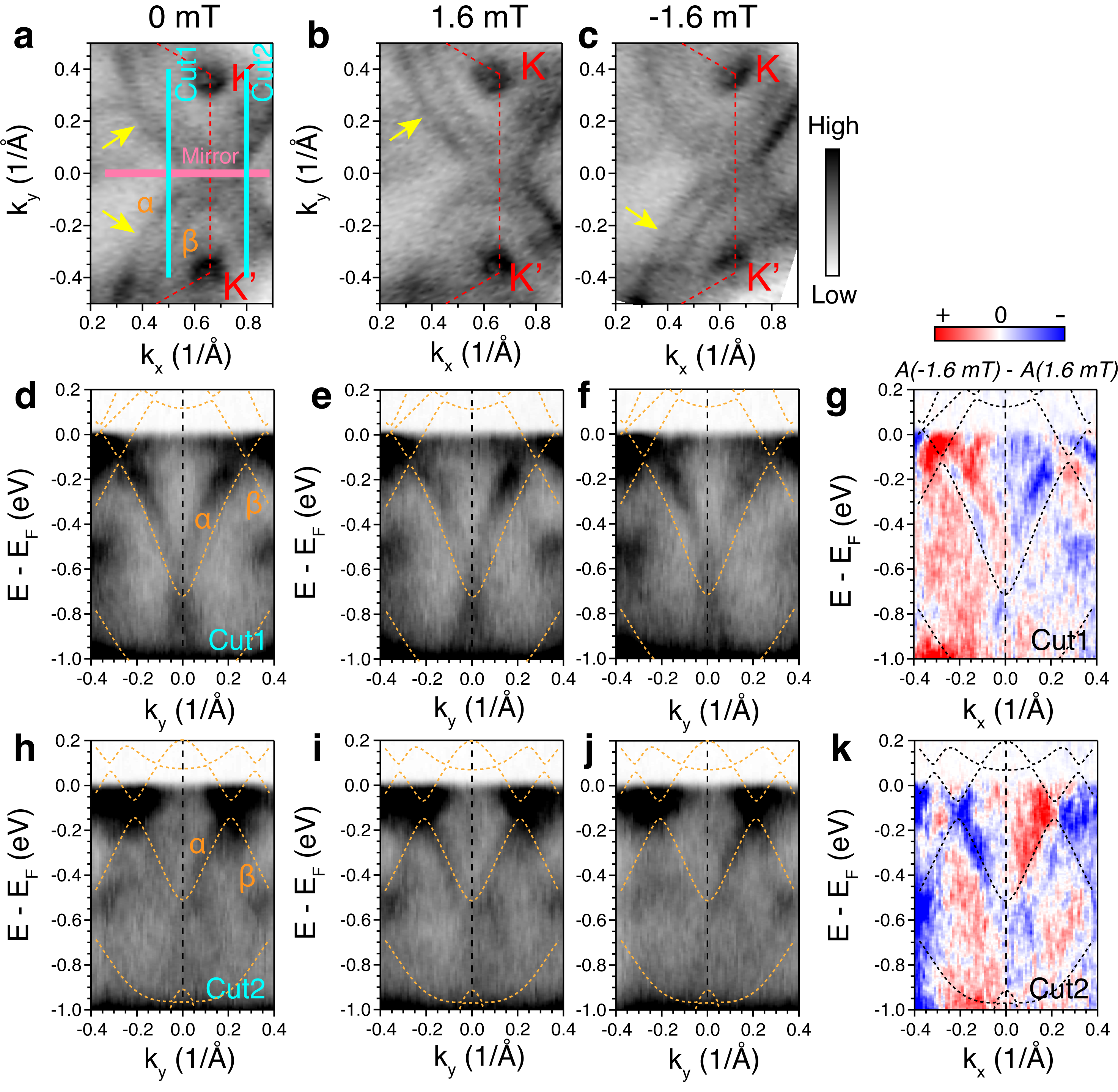}
\caption{\label{fig:FigS11} {\bf Electronic response of \cvs~around the K point under a magnetic field.} (a)-(c) Constant energy contours at -0.35 eV without a magnetic field, with +1.6 mT, and with -1.6 mT, respectively. (d)-(e) The corresponding spectral images along cut1 indicated in (a). (g) The spectral image obtained by subtracting (e) from (f). (h)-(k) Same as (d)-(g) but along cut2 indicated in (a).
}
\end{figure}

\begin{figure}
\includegraphics[width=0.85\textwidth]{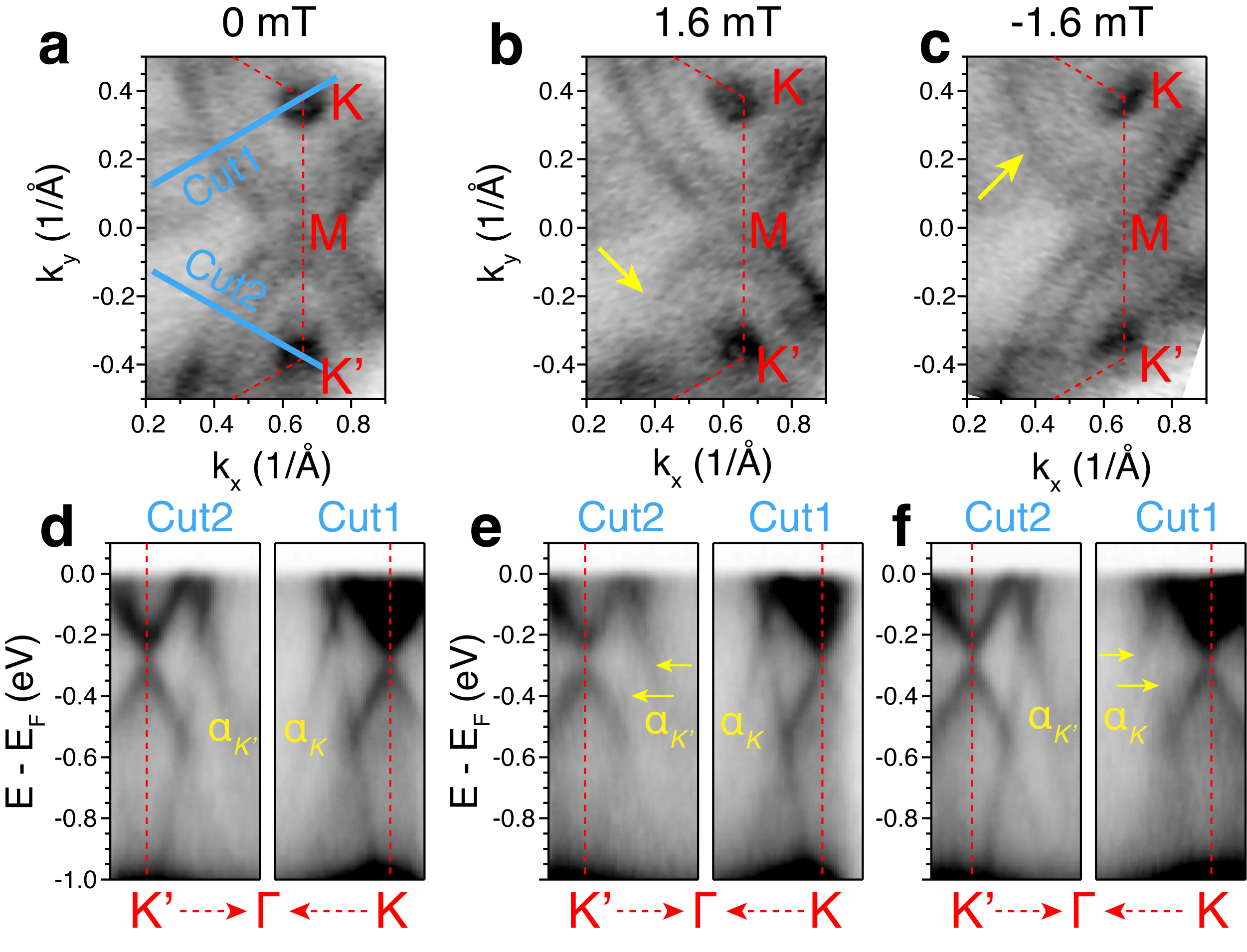}
\caption{\label{fig:FigS4} {\bf Momentum-selective spectral broadening in \cvs~under a magnetic field.} (a)-(c) Constant energy contours at -0.35 eV without a magnetic field, with +1.6 mT, and with -1.6 mT, respectively. The yellow arrows point to the Fermi surface sheets that broaden significantly under the corresponding magnetic field. (d)-(f) Spectral images taken along cut1 and cut2 indicated in (a), with cut directions perpendicular to the Fermi surface sheets. Selective momentum broadening of the $\alpha$ and $\beta$ bands is evident from both the Fermi surface sheets and the corresponding band spectral images.
}
\end{figure}

\begin{figure}
\includegraphics[width=0.98\textwidth]{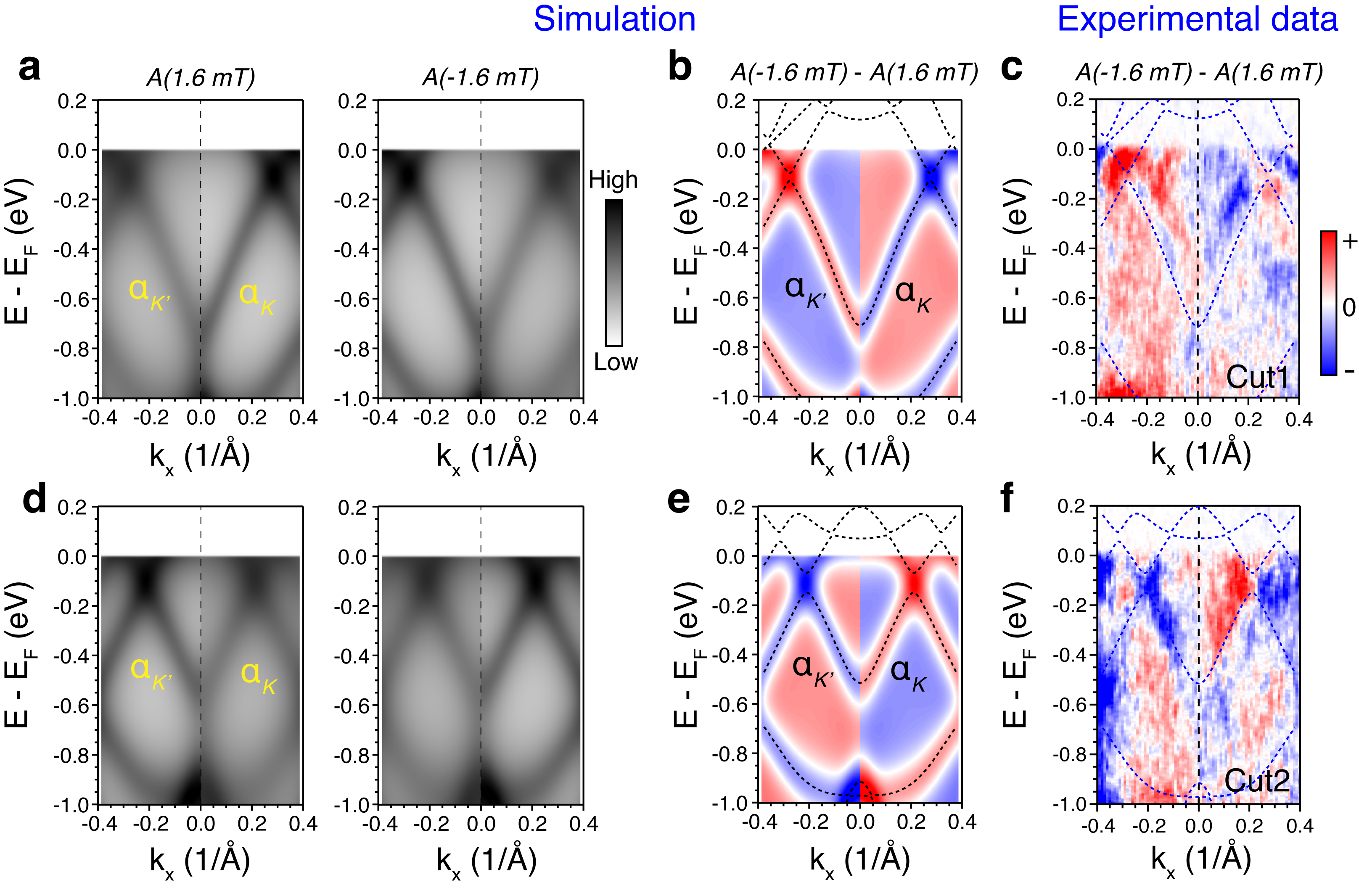}
\caption{\label{fig:FigS5} {\bf Simulation of the magneto-dichroic effect of the ARPES spectra.} (a) Simulated spectral images to reproduce the ARPES observations based on the band structure obtained by DFT calculations (black dashed line in (c)). The DFT band dispersions were extracted along the momentum path along cut1 in Fig. 2b of the main text. Distinct imaginary parts of the self-energies were applied to the left and right momentum regions in each panel to simulate the momentum-selective spectral broadening under a magnetic field. (b) Simulated magneto-dichroic spectral image by subtracting the left panel from the right panel in (a). (c) Measured magneto-dichroic spectral image from Fig. 2h in main text. (d)-(f) Same as (a)-(c) but simulating the magneto-dichroic spectral image along cut2 in Fig. 2b of the main text.
}
\end{figure}

\begin{figure}
\includegraphics[width=0.98\textwidth]{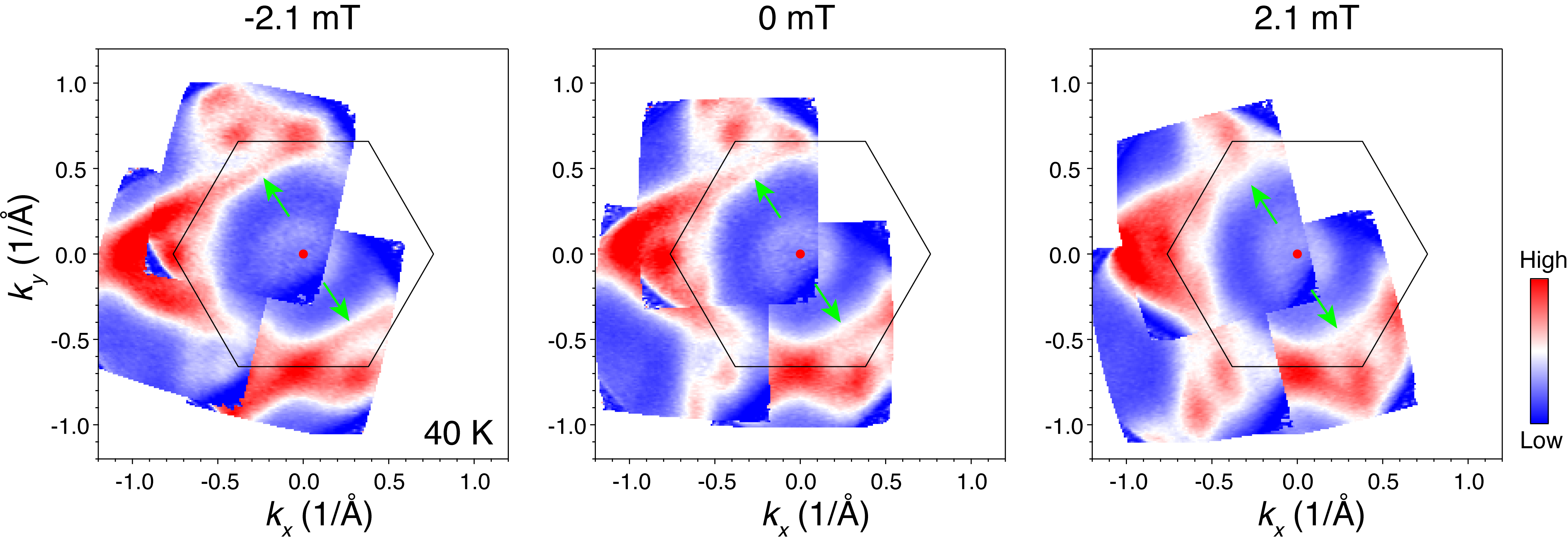}
\caption{\label{fig:FigS7} {\bf Constant energy contour mappings showing the two opposite K regions.} Constant energy contours at -0.2 eV consist of three separate DA30 deflector mode mappings on the same sample, showing the two opposite K regions using a helium lamp light source. The mappings were taken at magnetic fields of -2.1 mT, 0 mT, and 2.1 mT. The green arrows point to the Fermi surface sheets respond most prominently to the magnetic field. The two Fermi surface sheets around the two opposite K regions exhibit the same behavior, which is also evident by our observations in Fig. 2c and d of the main text considering the crystalline translational symmetry.
}
\end{figure}

\begin{figure}
\includegraphics[width=0.95\textwidth]{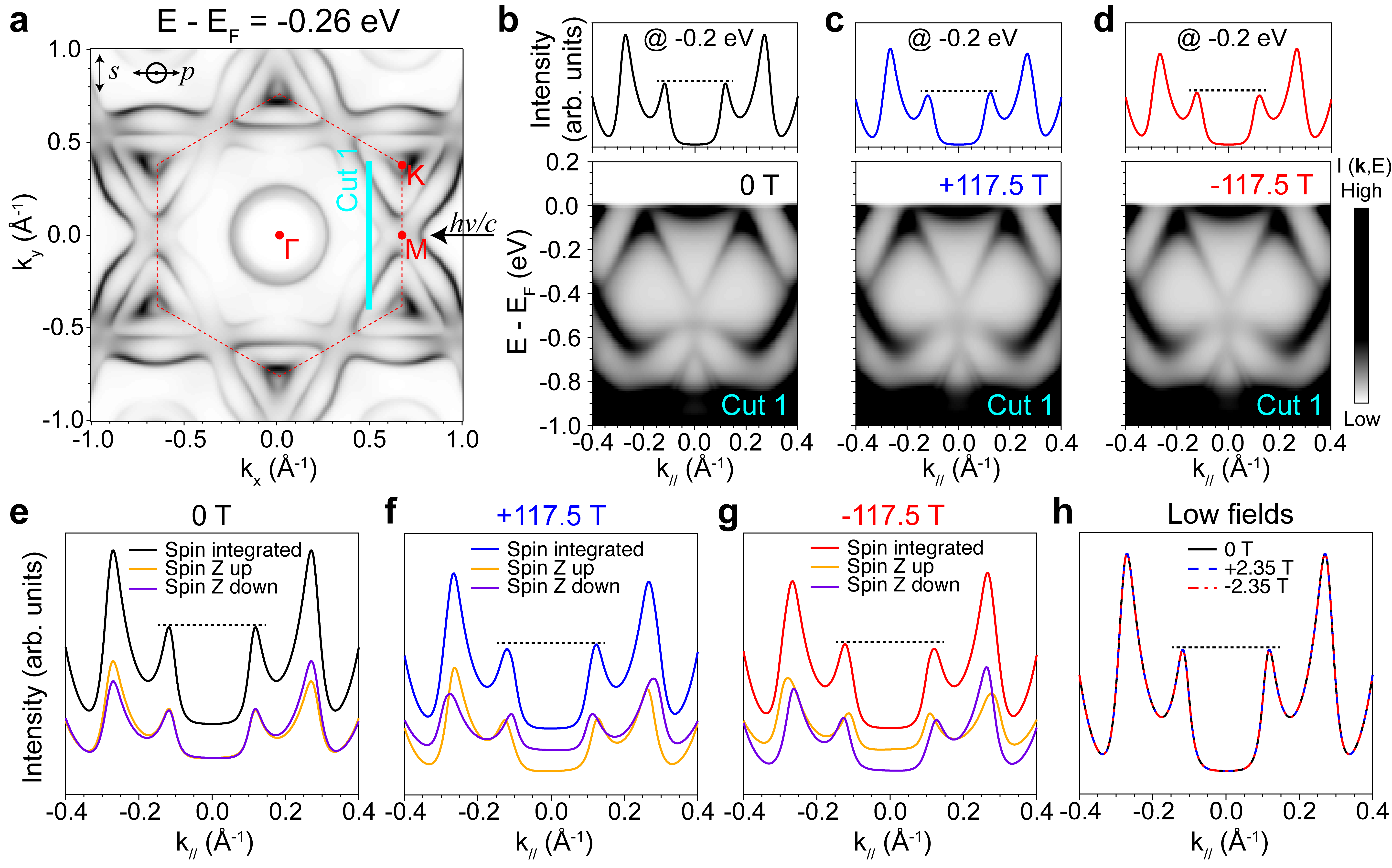}
\caption{\label{fig:Fig_onestep} \edt{{\bf One-step model ARPES calculations of \cvs~under external magnetic fields within the atomic-sphere approximation.} (a) Constant energy contour under zero field calculated at $E$ – \ef~= -0.26 eV using experimental geometry and inputs. Photon incidence direction and the $s$ and $p$ polarizations of photons are illustrated by the $h\nu/c$ and the symbols on the top left. (b) ARPES spectra of Cut 1 spanned by the vertical cyan line in (a), similar to experiments performed in main text Fig. 2(e). The momentum distribution curve (MDC) taken at $E$ – \ef = -0.2 eV is shown on top. A Fermi-Dirac distribution function of 20 K convolved with a putative 20 meV experimental resolution is applied to the calculated data. (c, d) Same as (b), but under external magnetic fields of +117.5 and -117.5 T, respectively. (e - g) Spin integrated MDCs reproduced from (b-d) and their spin-resolved components projected along z under 0, +117.5, and -117.5 T, respectively. (h) Spin-integrated MDCs taken at $E$ – \ef = -0.2 eV of Cut 1 but at low fields of 0, +2.35, and -2.35 T.
} }
\end{figure}

\begin{figure}
\includegraphics[width=0.98\textwidth]{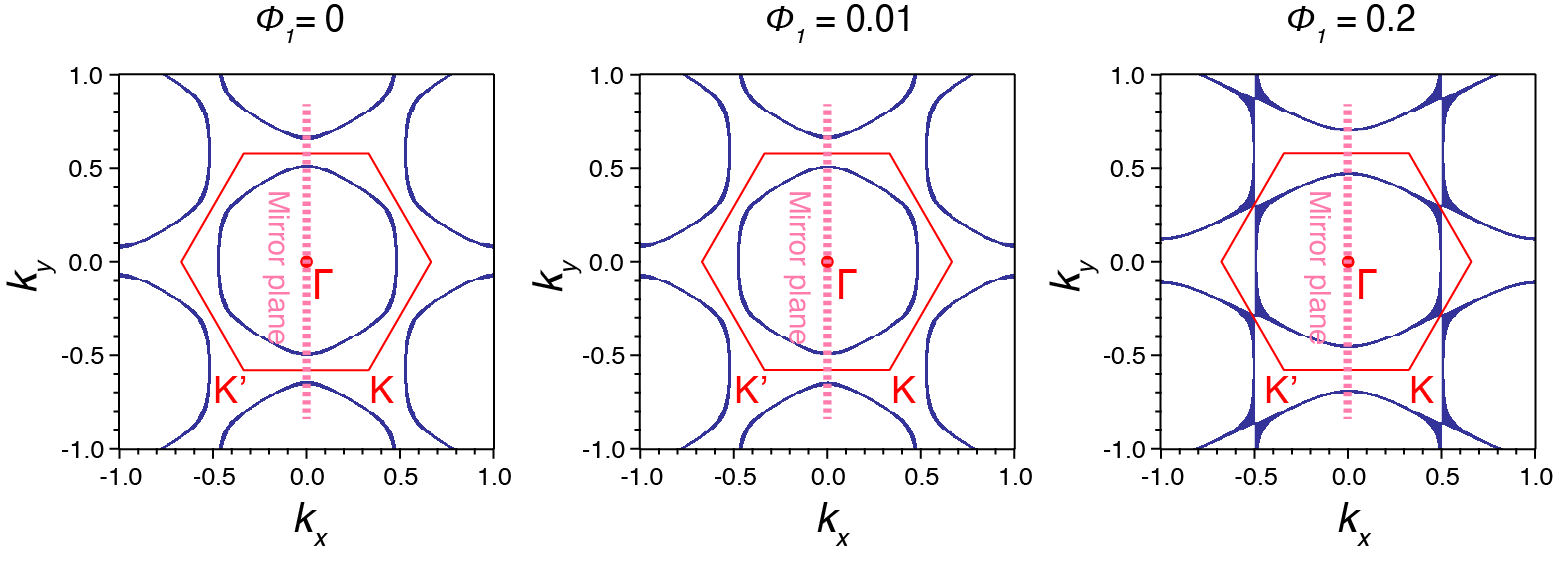}
\caption{\label{fig:FigS13} {\bf Fermi surface of the nearest-neighbor kagome model of Methods Section F in the absence of a magnetic field.} The parameters are set as $t = 1$ and $\mu = -0.08$. $B_z = 0$ to focus on the effect of $\Phi_1$, which is varied as indicated in each panel ($\Phi_1 = 0$, $0.01$, and $0.2$). As expected, the Fermi surface distortion is negligible for small $\Phi_1$. 
}
\end{figure}

\begin{figure}
\includegraphics[width=0.98\textwidth]{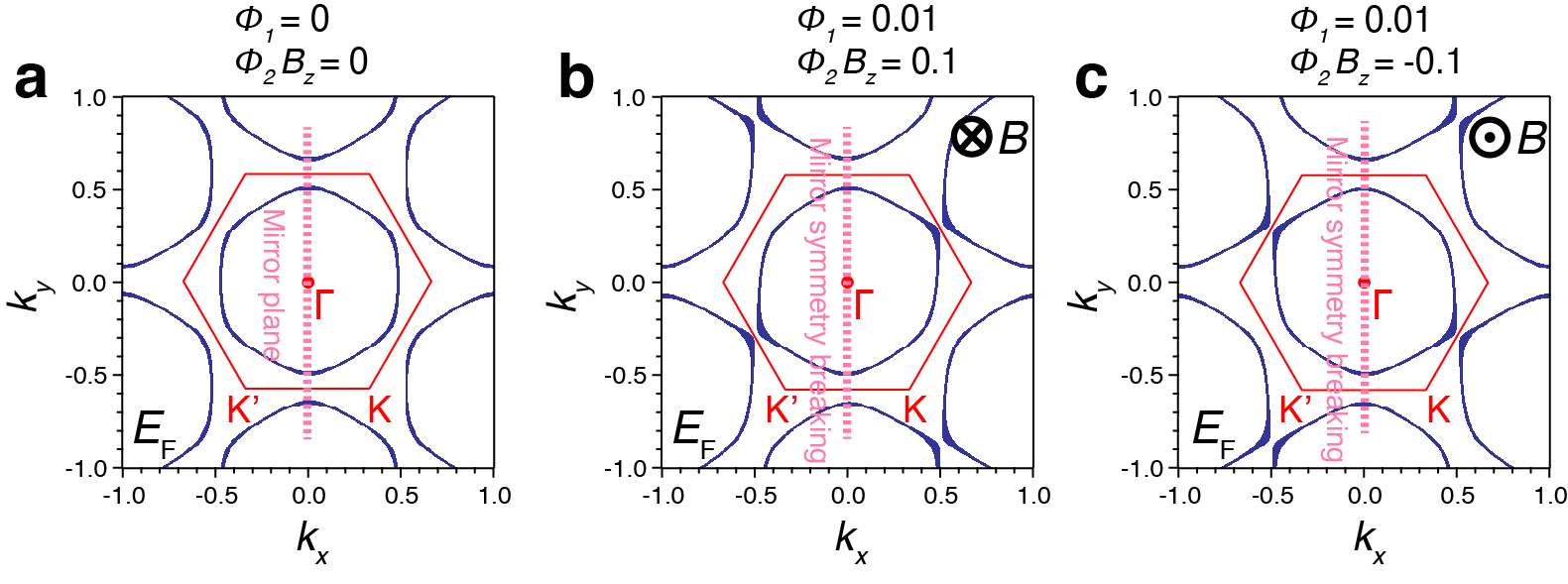}
\caption{\label{fig:FigS14} {\bf Fermi surface of the nearest-neighbor kagome model of Methods Section F in the presence of a magnetic field.} (a) Fermi surface for $t = 1$, $\mu = -0.08$, $\Phi_1 = 0$ and $B_z=0$. The horizontal mirror plane is preserved. (b) Fermi surface with $t = 1$ and $\mu = -0.08$, but now with $\Phi_1 = 0.01$ and a positive field $\Phi_2 B_z = 0.1$. (c) Same as (b) but with a negative field $\Phi_2B_z = -0.1$. In both (b) and (c), the horizontal mirror symmetry and the $C_6$ symmetries are broken, consistent with the symmetry-breaking behavior observed in the ARPES data from the perspective of the spectral weight distribution.
}
\end{figure}

\begin{figure}
\includegraphics[width=0.98\textwidth]{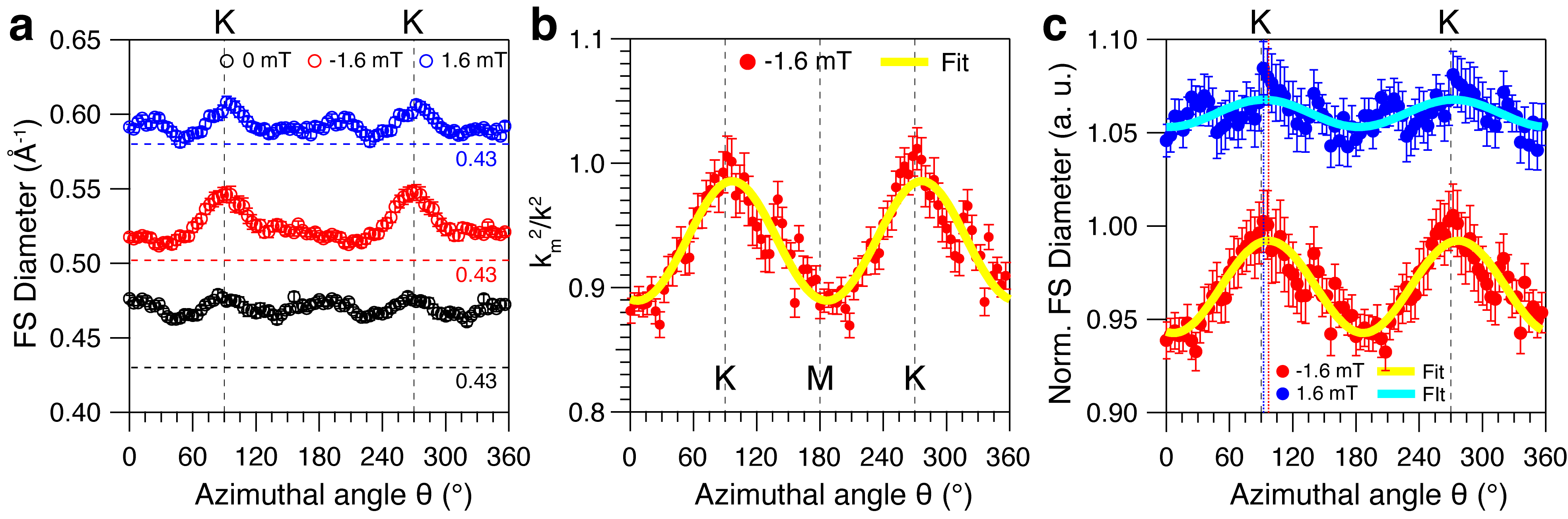}
\caption{\label{fig:FigS9_N2} {\bf \g-pocket Fermi surface fitting based on the model of Methods Section F, Eq. (\ref{equ:EqFit}).} (a) Fermi pocket diameter as a function of the azimuthal angle obtained by fitting the radial MDCs without and with a magnetic field. The curves at -1.6 mT and 1.6 mT are offset for clarity with the reference lines shown for each. (b) Square of the Fermi momentum for -1.6 mT normalized by the zero field value overlaid with the fitting results using Eq.~\ref{equ:EqFit}. (c) Fermi pocket diameter for -1.6 mT and 1.6 mT normalized by its zero-field value measured at 35 K. All curves are fitted with the function $A + B \cos(2(\theta+\phi))$. The red and blue dashed lines indicate the deviation of the long axes of the elliptical Fermi pockets away from the K point at -1.6 mT (6.9$^\circ$) and 1.6 mT (2.3$^\circ$), respectively.
}
\end{figure}

\end{document}